\documentclass{aa}
\usepackage{color}
\usepackage{graphicx}
\usepackage{lscape}
\usepackage{natbib}
\usepackage[scaled]{helvet}
\usepackage[varg]{txfonts}
\usepackage{url}
\usepackage{xspace}
\bibpunct{(}{)}{;}{a}{}{,}
\newcounter{Rco}
\newcommand{\ionw}[3]{\mbox{\ion{#1}{#2}~$\lambda\,#3\,\mathrm{\AA}$}\xspace}
\newcommand{\ionww}[3]{\mbox{\ion{#1}{#2}~$\lambda\lambda\,#3\,\mathrm{\AA}$}\xspace}
\newcommand{\Ionst}[1]{\setcounter{Rco}{#1}\Roman{Rco}}

\newcommand{\Ionw}[3]{\mbox{#1\,{\scriptsize\Ionst{#2}}~$\lambda\,#3$\,\AA}\xspace}

\newcommand{\Ionww}[3]{\mbox{#1\,{\scriptsize\Ionst{#2}}~$\lambda\lambda\,#3$\,\AA}\xspace}
\newcommand{\Jonw}[3]{\mbox{\ion{#1}{#2}~$\lambda\,#3$\,\AA}\xspace}

\newcommand{\loggw}[1]{\mbox{$\log g\hspace{-0.5mm} =\hspace{-0.5mm}  #1$}}

\newcommand{\Teff}{\mbox{$T_\mathrm{eff}$}\xspace}
\newcommand{\Teffw}[1]{\mbox{$\Teff\hspace{-0.5mm} =\hspace{-0.5mm} #1 \,\mathrm{K}$}}

\newcommand{\gb}{\object{G191$-$B2B}\xspace}
\newcommand{\re}{\object{RE\,0503$-$289}\xspace}
\begin{document}

\title{Stellar laboratories }
\subtitle{VIII. New \ion{Zr}{iv -- vii}, \ion{Xe}{iv -- v}, and \ion{Xe}{vii} oscillator strengths 
                and the Al, Zr, and Xe abundances in the hot white dwarfs \gb and \re
           \thanks
           {Based on observations with the NASA/ESA Hubble Space Telescope, obtained at the Space Telescope Science 
            Institute, which is operated by the Association of Universities for Research in Astronomy, Inc., under 
            NASA contract NAS5-26666.
           }\fnmsep
           \thanks
           {Based on observations made with the NASA-CNES-CSA Far Ultraviolet Spectroscopic Explorer.
           }\fnmsep
           \thanks
           {Tables \ref{tab:zriv:loggf} to \ref{tab:zrvii:loggf} and
                   \ref{tab:xeiv:loggf} to \ref{tab:xevii:loggf}      are only available via the
            German Astrophysical Virtual Observatory (GAVO) service TOSS (http://dc.g-vo.org/TOSS).
           }
         }

\titlerunning{Stellar laboratories: New \ion{Zr}{iv -- vii}, \ion{Xe}{iv -- v}, and \ion{Xe}{vii} oscillator strengths}

\author{T\@. Rauch\inst{1}
        \and
        S\@. Gamrath\inst{2}
        \and
        P\@. Quinet\inst{2,3}
        \and
        L\@. L\"obling\inst{1}
        \and
        D\@. Hoyer\inst{1}
        \and
        K\@. Werner\inst{1}
        \and
        J\@. W\@. Kruk\inst{4}
        \and
        M\@. Demleitner\inst{5}
        }

\institute{Institute for Astronomy and Astrophysics,
           Kepler Center for Astro and Particle Physics,
           Eberhard Karls University,
           Sand 1,
           72076 T\"ubingen,
           Germany \\
           \email{rauch@astro.uni-tuebingen.de}
           \and
           Physique Atomique et Astrophysique, Universit\'e de Mons -- UMONS, 7000 Mons, Belgium
           \and
           IPNAS, Universit\'e de Li\`ege, Sart Tilman, 4000 Li\`ege, Belgium
           \and
           NASA Goddard Space Flight Center, Greenbelt, MD\,20771, USA
           \and
           Astronomisches Rechen-Institut (ARI), Centre for Astronomy of Heidelberg University, M\"onchhofstra\ss e 12-14, 69120 Heidelberg, Germany}

\date{Received 27 September 2016; accepted 6 November 2016}

\abstract {For the spectral analysis of high-resolution and high-signal-to-noise spectra of hot stars,
           state-of-the-art non-local thermodynamic equilibrium (NLTE) 
           model atmospheres are mandatory. These are strongly
           dependent on the reliability of the atomic data that is used for their calculation.
          }
          {To search for zirconium and xenon lines in the ultraviolet (UV) spectra of \gb and \re,
           new \ion{Zr}{iv-vii}, \ion{Xe}{iv-v}, and \ion{Xe}{vii} oscillator strengths were calculated.
           This allows, for the first time, determination of the Zr abundance in white dwarf (WD) stars and
           improvement of the Xe abundance determinations.
          }
          {We calculated \ion{Zr}{iv-vii}, \ion{Xe}{iv-v}, and \ion{Xe}{vii} oscillator strengths
           to consider radiative and collisional bound-bound transitions of Zr and Xe
           in our NLTE stellar-atmosphere models for the analysis of their lines exhibited 
           in UV observations of the hot WDs \gb and \re.
          }
          {We identified 
           one new \ion{Zr}{iv}, 
            14 \ion{new Zr}{v}, and
            ten \ion{new Zr}{vi} 
           lines in the spectrum of \re.
           Zr was detected for the first time in a WD.
           We measured a Zr abundance of $-3.5 \pm 0.2$ (logarithmic mass fraction, approx. 11\,500 times solar).
           We identified five \ion{new Xe}{vi} lines and determined a Xe abundance of $-3.9 \pm 0.2$ (approx. 7\,500 times solar).
           We determined a preliminary photospheric Al abundance of $-4.3 \pm 0.2$ (solar) in \re.
           In the spectra of \gb, no Zr line was identified. The strongest \ion{Zr}{iv} line (1598.948\,\AA) in 
           our model gave an upper limit of $-5.6 \pm 0.3$ (approx. 100 times solar).
           No Xe line was identified in the UV spectrum of \gb and we confirmed the previously determined
           upper limit of $-6.8 \pm 0.3$ (ten times solar).
          }
          {Precise measurements and calculations of atomic data are a prerequisite for
           advanced NLTE stellar-atmosphere modeling. 
           Observed \ion{Zr}{iv-vi} and \ion{Xe}{vi-vii} line profiles in the UV spectrum of \re 
           were simultaneously well reproduced with our newly calculated oscillator strengths.
          }

\keywords{atomic data --
          line: identification --
          stars: abundances --
          stars: individual: \gb\ --
          stars: individual: \re\ --
          virtual observatory tools
         }

\maketitle

\section{Introduction}
\label{sect:intro}

The DO-type white dwarf (WD) star \re \citep[\object{WD\,0501+527},][]{mccooksion1999,mccooksion1999cat},
exhibits many lines of the trans-iron elements 
Zn (atomic number $Z = 30$), 
Ga (31), 
Ge (32), 
As (33), 
Se (34), 
Kr (36), 
Mo (42),
Sn (50), 
Te (52), 
I  (53),
Xe (54), and
Ba (56) 
in its ultraviolet spectrum. These were initially identified by \citet{werneretal2012}, who
determined the Kr and Xe abundances (Sect.\,\ref{sect:results}) based on atomic data available at that time. 
Calculations of transition
probabilities for Zn, Ga, Ge, Kr, Mo, Xe, and Ba in the subsequent years allowed precise abundance 
measurements for these elements 
\citep[][respectively]{
rauchetal2014zn, 
rauchetal2015ga, 
rauchetal2012ge, 
rauchetal2016mo, 
rauchetal2014ba,
rauchetal2015xe,
rauchetal2016kr}. 

Here we report that we have identified lines of an additional element, namely zirconium (40) which has never been detected 
before in WDs, and calculated new \ion{Zr}{iv-vii} 
transition probabilities to determine its photospheric abundance. To verify the Xe abundance determination
of \citet{werneretal2012}, we calculated much more complete \ion{Xe}{iv-v} and \ion{Xe}{vi} transition probabilities.

The hot, hydrogen-rich, DA-type WD \gb \citep[\object{WD\,0501+527},][]{mccooksion1999,mccooksion1999cat} is 
a primary flux reference standard for all absolute calibrations from 1000 to 25\,000\,\AA\ \citep{bohlin2007}.
\citet{rauchetal2013} presented a detailed spectral analysis of this star.
Based on their model, \citet{rauchetal2014zn,rauchetal2015ga,rauchetal2014ba} identified Zn, Ga, and Ba
lines in the observed UV spectrum and determined the abundances of these elements.

We briefly introduce our observational data in Sect.\,\ref{sect:observation}. The discovery of the
interstellar \ionww{Mg}{ii}{2796.35,2803.53} resonance doublet and its modelling is shown in
Sect.\,\ref{sect:ism}. Our model atmospheres are described in Sect.\,\ref{sect:models}. 
We start our spectral analysis with a search for Al lines and an abundance determination in Sect.\,\ref{sect:al}.
The Zr transition-probability calculation, line identification, and abundance analysis
are presented in Sect.\,\ref{sect:zr}, followed by the same for Xe in Sect.\,\ref{sect:xe}. 
We summarize our results and conclude in Sect.\,\ref{sect:results}.

\section{Observations}
\label{sect:observation}

For \re, we analyzed ultraviolet (UV) observations that were obtained with the
Far Ultraviolet Spectroscopic Explorer 
(FUSE, $910\,\mathrm{\AA} < \lambda <  1188\,\mathrm{\AA}$, resolving power $R = \lambda/\Delta\lambda \approx 20\,000$) 
and the
Hubble Space Telescope / Space Telescope Imaging Spectrograph 
(HST/STIS, $1144\,\mathrm{\AA} < \lambda < 3073\,\mathrm{\AA}$, $R \approx 45\,800$).
These were described in detail by
\citet{werneretal2012} and \citet{rauchetal2016kr}, respectively.

For \gb, we used the FUSE observation described by \citet{rauchetal2013}
and
the high-dispersion \'echelle spectrum 
\citep[HST/STIS, $1145\,-\,3145$\,\AA, $R \approx 100\,000$,][]{rauchetal2013}  available
from the CALSPEC\footnote{\url{http://www.stsci.edu/hst/observatory/cdbs/calspec.html}} database.

To compare observations with synthetic spectra, the latter were convolved with Gaussians to model the respective resolving power. 
The observed spectra are shifted to rest wavelengths according to radial-velocity measurements of 
$v_\mathrm{rad} = 24.56\,\mathrm{km\,s^{-1}}$ \citep{lemoineetal2002} and 
                $25.8\,\mathrm{km\,s^{-1}}$ for \gb and \re (our value), respectively.

\section{Interstellar line absorption}
\label{sect:ism}

\begin{figure}
   \resizebox{\hsize}{!}{\includegraphics{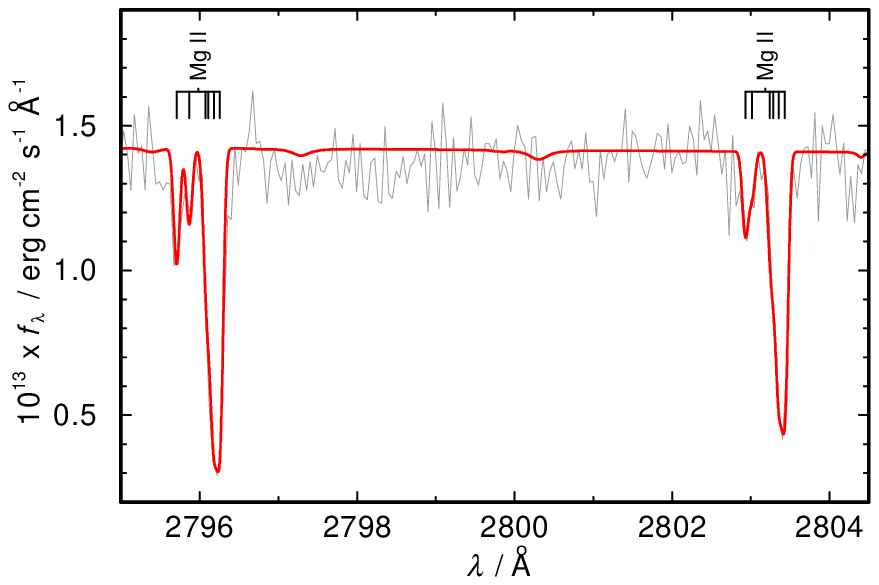}}
    \caption{Section of the STIS spectrum of \re with the interstellar \ionww{Mg}{ii}{2796.35,2803.53} lines.
            }
   \label{fig:ism_mg}
\end{figure}

\citet{rauchetal2016kr} found that the interstellar line absorption toward \re has a multi-velocity structure (radial-velocities
$-40\,\mathrm{km/s} < v_\mathrm{rad} < +18\,\mathrm{km/s}$). In the HST/STIS spectra of \re, the
interstellar \ionww{Mg}{ii}{2796.35,2803.53} resonance lines 
(3s\,$^2$S$_{\mathrm{1/2}}$ - 3p\,$^2$P$^\mathrm{o}_{\mathrm{3/2}}$ and
 3s\,$^2$S$_{\mathrm{1/2}}$ - 3p\,$^2$P$^\mathrm{o}_{\mathrm{1/2}}$ with oscillator strengths of
0.608 and 0.303, respectively)
are prominent (Fig.\,\ref{fig:ism_mg}) and corroborate
such a structure. Table\,\ref{tab:is} displays the parameters that were used to fit the observation.

\begin{table}\centering
\caption{Column densities (in $\mathrm{cm^{-2}}$) and radial velocities (in $\mathrm{km/s}$) 
         used to model interstellar clouds in the line of sight toward \re.}         
\label{tab:is}
\setlength{\tabcolsep}{.16em}
\begin{tabular}{cr@{.}lccr@{.}l}
\hline
\hline
\noalign{\smallskip}
\multicolumn{3}{c}{\ionw{Mg}{ii}{2796.35}} && 
\multicolumn{3}{c}{\ionw{Mg}{ii}{2803.53}} \\ 
\cline{1-3}
\cline{5-7}
\noalign{\smallskip}
$N$ & \multicolumn{2}{c}{$v_\mathrm{rad}$} && 
$N$ & \multicolumn{2}{c}{$v_\mathrm{rad}$} \\
\cline{1-3}
\cline{5-7}
\noalign{\smallskip}
$2.9 \times 10^{12}$ & $+15$&$0$ &&  $4.5 \times 10^{12}$ & $+15$&$0$ \\
$2.6 \times 10^{12}$ & $ +7$&$0$ &&  $3.8 \times 10^{12}$ & $ +7$&$0$ \\
$8.0 \times 10^{11}$ & $ -0$&$5$ &&  $1.2 \times 10^{12}$ & $ -0$&$5$ \\
$4.6 \times 10^{11}$ & $ -4$&$5$ &&  $8.5 \times 10^{11}$ & $ -5$&$5$ \\
$4.5 \times 10^{11}$ & $-26$&$5$ &&  $5.0 \times 10^{11}$ & $-29$&$5$ \\
$7.3 \times 10^{11}$ & $-43$&$5$ &&  $1.0 \times 10^{12}$ & $-38$&$5$ \\
\hline
\end{tabular}
\end{table}

\section{Model atmospheres and atomic data}
\label{sect:models}

We calculated plane-parallel, chemically homogeneous model-atmospheres in hydrostatic and radiative 
equilibrium with the T\"ubingen non-local thermodynamic equilibrium (NLTE) Model Atmosphere Package
\citep[TMAP\footnote{\url{http://astro.uni-tuebingen.de/~TMAP}},][]{werneretal2003,tmap2012}.
Model atoms were retrieved from the T\"ubingen Model Atom Database
\citep[TMAD\footnote{\url{http://astro.uni-tuebingen.de/~TMAD}},][]{rauchdeetjen2003} that
has been constructed as part of the T\"ubingen contribution to the German Astrophysical Virtual Observatory 
(GAVO\footnote{\url{http://www.g-vo.org}}).

The effective temperatures, surface gravities, and photospheric abundances 
of
\gb \citep[\Teffw{60\,000 \pm 2000}, $\log\,(g\,/\,\mathrm{cm\,s^{-2}}) = 7.6 \pm 0.05$,][]{rauchetal2013}
and 
\re \citep[\Teffw{70\,000 \pm 2000}, \loggw{7.50 \pm 0.1},][]{rauchetal2016kr}
were previously analyzed with TMAP models. We adopt these parameters for our calculations.

\ion{Zr}{iv-vii} and \ion{Xe}{iv-vii} were represented by the Zr and Xe model atoms
with so-called super levels and super lines that were calculated with a statistical approach via our 
Iron Opacity and Interface 
\citep[IrOnIc\footnote{\url{http://astro.uni-tuebingen.de/~TIRO}},][]{rauchdeetjen2003,muellerringatPhD2013}.
To enable IrOnIc to read our new Zr and Xe data, we transferred it into Kurucz-formatted 
files \citep[cf.,][]{rauchetal2015ga}. 
The statistics of our Zr and Xe model atoms is listed in Table\,\ref{tab:ironic}.

\begin{table}\centering
\caption{Statistics of \ion{Zr}{iv - vii} and \ion{Xe}{iv - v, vii} atomic levels and line transitions from
         Tables\,\ref{tab:zriv:loggf} - \ref{tab:zrvii:loggf} and
                 \ref{tab:xeiv:loggf} - \ref{tab:xevii:loggf}, respectively.
         \ion{Xe}{vi} is shown for completeness.
        } 
\label{tab:ironic}
\begin{tabular}{lcccc}
\hline
\hline
ion                             & atomic levels & lines & super levels & super lines \\
\hline
\ion{Zr}{iv}                    &            52 &   135 &            7 &          20 \\
\ion{Zr}{v}                     &           135 &  1449 &            7 &          22 \\
\ion{Zr}{vi}                    &            96 &  1098 &            7 &          12 \\
\ion{Zr}{vii}                   &            83 &   947 &            7 &          15 \\
\hline
\multicolumn{1}{r}{total}       &           366 &  3629 &           28 &          69 \vspace{3mm}\\
\hline
\ion{Xe}{iv}                    &            94 &  1391 &            7 &          16 \\
\ion{Xe}{v}                     &            65 &   616 &            7 &          15 \\
\ion{Xe}{vi}\tablefootmark{a}   &            90 &   243 &            7 &          16 \\
\ion{Xe}{vii}                   &            60 &   491 &            7 &          19 \\
\hline                                                             
\multicolumn{1}{r}{total}       &           309 &  2741 &           28 &          66 \vspace{1mm}\\
\hline
\end{tabular}
\tablefoot{
\tablefoottext{a}{Atomic level and line data taken from \citet{gallardoetal2015}.}
}
\end{table}

For Zr and Xe and all other species, level dissolution (pressure ionization) following
\citet{hummermihalas1988} and \citet{hubenyetal1994} is accounted for. 
Broadening for all Al, Zr, and Xe lines due to the quadratic Stark effect is calculated 
using approximate formulae given by \citet{cowley1970,cowley1971}.

All spectral energy distributions (SEDs) that were calculated for this analysis are available via
the registered Theoretical Stellar Spectra Access 
(TheoSSA\footnote{\url{http://dc.g-vo.org/theossa}}) GAVO service.

\section{Aluminum in \re}
\label{sect:al}

\begin{figure}
   \resizebox{\hsize}{!}{\includegraphics{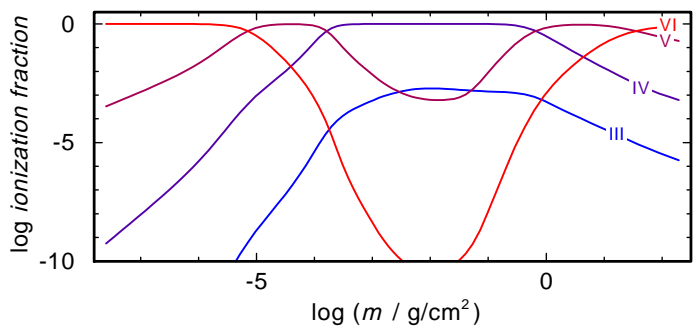}}
    \caption{Al ionization fractions in our \gb model.
             $m$ is the column mass, measured from the outer boundary of our model atmospheres.
            }
   \label{fig:ion_al_gb}
\end{figure}

\begin{figure}
   \resizebox{\hsize}{!}{\includegraphics{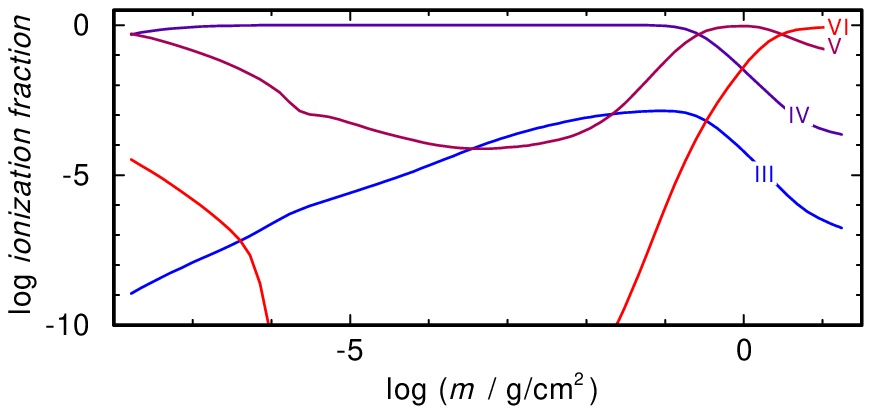}}
    \caption{As Fig.\,\ref{fig:ion_al_gb}, for \re.
            }
   \label{fig:ion_al_re}
\end{figure}

The Al abundance in \re was hitherto undetermined.
TMAD provides a recently extended Al model atom (Table\,\ref{tab:al}). We used it
to search for Al lines in the UV and optical spectra of \gb and \re,
especially for \ion{Al}{iv} lines, because, in both stars, this is  the dominant ionization stage in the line-forming region
($-4 \la \log\,m \la 0.5$, Figs.\,\ref{fig:ion_al_gb}, \ref{fig:ion_al_re}). So far, only \ion{Al}{iii} lines were identified 
in the UV spectrum of \gb, namely
$\lambda\lambda 1854.714, 1862.787\,\mathrm{\AA}$ \citep{holbergetal1998} 
and
$\lambda\lambda 1379.668, 1384.130, 1605.764, 1611.812, 1611.854 \,\mathrm{\AA}$
\citep[][logarithmic mass fraction of Al\,$ = -4.95 \pm 0.2$]{rauchetal2013}.

\begin{table}\centering
\caption{Statistics of the Al model atom used in our calculations compared
         to our previous analyses \citep[e.g.,][]{rauchetal2013, rauchetal2016kr}.}         
\label{tab:al}
\begin{tabular}{lccccc}
\hline
\hline        & \multicolumn{2}{c}{this work} &~~& \multicolumn{2}{c}{previous analyses} \\
\cline{2-3}
\cline{5-6}
              &               &       &&               &       \vspace{-6.0mm}\\
ion           &               &       &&               &       \vspace{-1.5mm}\\
              & atomic levels & lines && atomic levels & lines \\
\hline
\ion{Al}{ii } &               &       &&             1 &     0 \\
\ion{Al}{iii} &            24 &    70 &&             7 &    10 \\
\ion{Al}{iv}  &            61 &   276 &&             6 &     3 \\
\ion{Al}{v}   &            43 &   168 &&             6 &     4 \\
\ion{Al}{vi}  &             1 &     0 &&             1 &     0 \\
\hline                                                       
              &           129 &   514 &&            21 &    17 \\
\hline
\end{tabular}
\end{table}

\begin{figure*}
   \resizebox{\hsize}{!}{\includegraphics{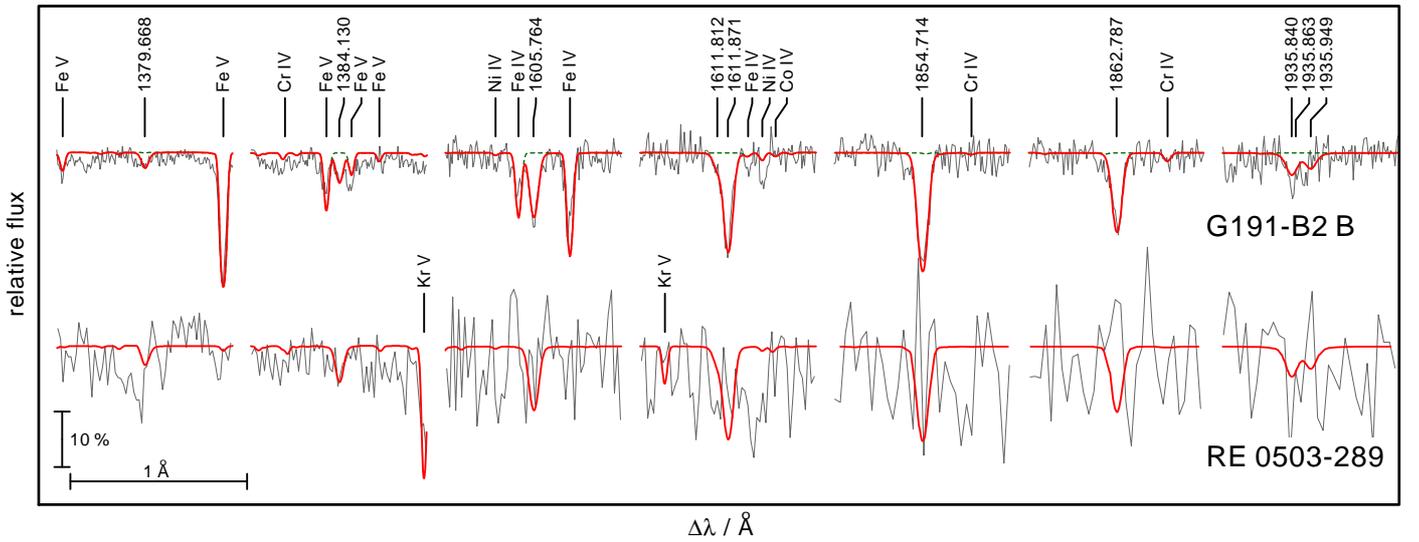}}
    \caption{Comparison of sections of the STIS spectra with our models for
             \gb (top) and \re (bottom). The Al abundances are
             $1.1 \times 10^{-5}$ \citep[0.2 times the solar value, ][]{rauchetal2013} and 
             $5.3 \times 10^{-5}$ (solar), respectively.
             In the top panel, the green dashed line is a spectrum calculated without Al.
             Prominent lines are marked, the identified \ion{Al}{iii} lines with their
             wavelengths.
            }
   \label{fig:al}
\end{figure*}

The only additional Al lines found in the observed spectra of \gb are
\Ionww{Al}{3}{1935.840, 1935.863,$ and $  1935.949} (Fig.\,\ref{fig:al}). \ion{Al}{iv} lines in our model are
entirely too weak to detect them in the observations.
Compared to the available STIS spectrum of \gb, that of \re has a much lower signal-to-noise ratio (S/N) that
hampers detection of Al lines. \Ionww{Al}{3}{1384.130} is the only line that is present in the
observation and is well reproduced at a solar Al abundance ($-4.28 \pm 0.2$). 
This result is based on a single line only, and thus it must be judged as uncertain. It is, however,
at least an upper abundance limit. The derived abundance is, nonetheless, in good agreement with the expectation
(interpolation in Fig.\,\ref{fig:X}). To improve the Al abundance measurement, better UV spectra for \re 
are highly desirable.

\section{Zirconium}
\label{sect:zr}

\begin{figure}
   \resizebox{\hsize}{!}{\includegraphics{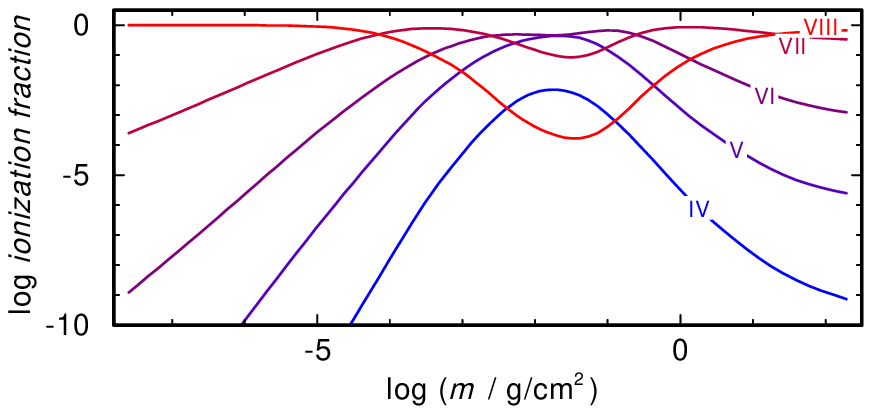}}
    \caption{Like Fig.\,\ref{fig:ion_al_gb}, for Zr.
            }
   \label{fig:ion_zr_gb}
\end{figure}

\begin{figure}
   \resizebox{\hsize}{!}{\includegraphics{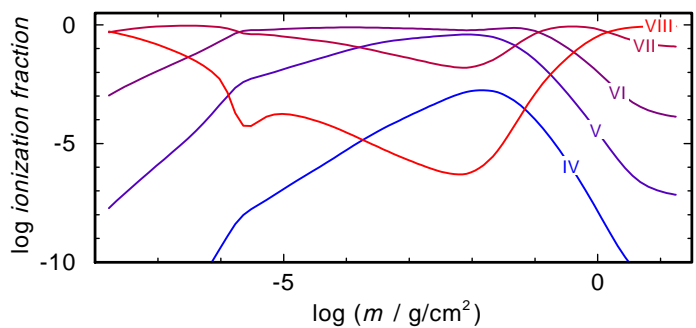}}
    \caption{Like Fig.\,\ref{fig:ion_al_re}, for Zr.
            }
   \label{fig:ion_zr_re}
\end{figure}

\subsection{Oscillator-strength calculations for Zr {\small IV} - {\small VII} ions}
\label{sect:zrtrans}

Radiative decay rates (oscillator strengths and transition probabilities) were computed using the 
pseudo-relativistic Hartree-Fock (HFR) method originally introduced by \citet{cowan1981}, and modified for 
taking into account core-polarization effects (CPOL), giving rise to the HFR+CPOL approach 
\citep[e.g.,][]{quinetetal1999,quinetetal2002}. 

For \ion{Zr}{iv}, configuration interaction was considered among the configurations 
4s$^2$4p$^6$$n$d ($n$ = 4 -- 9), 
4s$^2$4p$^6$$n$s ($n$ = 5 -- 9), 
4s$^2$4p$^6$$n$g ($n$ = 5 -- 9), 
4s$^2$4p$^6$$n$i ($n$ = 7 -- 9), 
4s$^2$4p$^5$4d5p, 
4s$^2$4p$^5$4d4f, and 
4s$^2$4p$^5$4d5f for the even parity, and 
4s$^2$4p$^6$$n$p ($n$ = 5 -- 9), 
4s$^2$4p$^6$$n$f ($n$ = 4 -- 9), 
4s$^2$4p$^6$$n$h ($n$ = 6 -- 9), 
4s$^2$4p$^6$$n$k ($n$ = 8 -- 9), 
4s$^2$4p$^5$4d$^2$, 
4s$^2$4p$^5$4d5s, and 
4s$^2$4p$^5$4d5d for the odd parity. The core-polarization parameters were the dipole polarizability of a \ion{Zr}{vi} ionic 
core as reported by \citet{fragaetal1976}, that is, $\alpha_\mathrm{d}$ = 2.50\,a.u., and the cut-off radius corresponding to the 
HFR mean value $<$$r$$>$ of the outermost core orbital (4p), that is, $r_c$ = 1.34\,a.u\@. Using the experimental energy levels 
taken from the analysis by \citet{readeracquista1997}, the average energies and spin-orbit parameters of 
4s$^2$4p$^6$$n$d ($n$ = 4 -- 6), 
4s$^2$4p$^6$$n$s ($n$ = 5 -- 8), 
4s$^2$4p$^6$$n$g ($n$ = 5 -- 9), 
4s$^2$4p$^6$$n$p ($n$ = 5 -- 7), 
4s$^2$4p$^6$$n$f ($n$ = 4 -- 6), and 
4s$^2$4p$^6$6h configurations were adjusted using a well-established least-squares fitting procedure in which the 
mean deviations with experimental data were found to be equal to 0\,cm$^{-1}$ for the even parity and 6\,cm$^{-1}$ for 
the odd parity. 

For \ion{Zr}{v}, the configurations explicitly included in the HFR model were 
4s$^2$4p$^6$, 
4s$^2$4p$^5$$n$p ($n$ = 5 -- 7), 
4s$^2$4p$^5$$n$f ($n$ = 4 -- 7), 
4s4p$^6$$n$d ($n$ = 4 -- 7), 
4s4p$^6$$n$s ($n$ = 5 -- 7), 
4s$^2$4p$^4$4d$^2$, 
4s$^2$4p$^4$4d5s, and 
4s$^2$4p$^4$5s$^2$ for the even parity, and 
4s$^2$4p$^5$$n$d ($n$ = 4 -- 7), 
4s$^2$4p$^5$$n$s ($n$ = 5 -- 10), 
4s$^2$4p$^5$$n$g ($n$ = 5 -- 7), 
4s4p$^6$$n$p ($n$ = 5 -- 7), 
4s4p$^6$$n$f ($n$ = 4 -- 7), 
4s$^2$4p$^4$4d5p, and 
4s$^2$4p$^4$4d4f for the odd parity. Core-polarization effects were estimated using $\alpha$$_d$ = 0.08\,a.u\@. and 
$r_c$ = 0.45\,a.u\@. These values correspond to a Ni-like \ion{Zr}{xiii} ionic core, with 3d as an outermost core subshell. 
In this ion, the semi-empirical process was performed to optimize the average energies, spin-orbit parameters, and 
electrostatic interaction Slater integrals corresponding to 
4p$^6$, 
4p$^5$$n$p ($n$ = 5 -- 6), 
4p$^5$4f, 
4s4p$^6$4d, 
4p$^5$$n$d ($n$ = 4 -- 7), 
4p$^5$$n$s ($n$ = 5 -- 10), 
4p$^5$$n$g ($n$ = 5 -- 6), and 
4s4p$^6$5p configurations using the experimental levels reported by \citet{readeracquista1979} and \citet{khanetal1981}. 
The mean deviations between calculated and experimental energies were 77\,cm$^{-1}$ and 91\,cm$^{-1}$ for even and 
odd parities, respectively.

In the case of \ion{Zr}{vi}, the HFR method was used with the interacting configurations 
4s$^2$4p$^5$, 
4s$^2$4p$^4$$n$p ($n$ = 5 -- 6), 
4s$^2$4p$^4$$n$f ($n$ = 4 -- 6), 
4s4p$^5$$n$d ($n$ = 4 -- 6), 
4s4p$^5$$n$s ($n$ = 5 -- 6), 
4p$^6$$n$p ($n$ = 5 -- 6), 
4p$^6$$n$f ($n$ = 4 -- 6), 
4s$^2$4p$^3$4d$^2$, 
4s$^2$4p$^3$4d5s, and 
4s$^2$4p$^3$5s$^2$ for the odd parity, and 
4s4p$^6$, 
4s$^2$4p$^4$$n$d ($n$ = 4 -- 6), 
4s$^2$4p$^4$$n$s ($n$ = 5 -- 6), 
4s$^2$4p$^4$$n$g ($n$ = 5 -- 6), 
4s4p$^5$$n$p ($n$ = 5 -- 6), 
4s4p$^5$$n$f ($n$ = 4 -- 6), 
4p$^6$$n$s ($n$ = 5 -- 6), 
4p$^6$$n$d ($n$ = 4 -- 6), 
4s$^2$4p$^3$4d5p, and 
4s$^2$4p$^3$4d4f for the even parity. Core-polarization effects were estimated using the same $\alpha$$_d$ and $r_c$ values 
as those considered in \ion{Zr}{v}. The radial integrals corresponding to 
4p$^5$, 
4p$^4$5p, 
4s4p$^6$, 
4p$^4$5d, 
4p$^4$5s, and 
4p$^4$6s were adjusted to minimize the differences between the calculated Hamiltonian eigenvalues and the experimental energy 
levels taken from \citet{readerlindsay2016}. In this process, we found mean deviations equal to 111\,cm$^{-1}$ in the odd 
parity and 221\,cm$^{-1}$ in the even parity.

Finally, for \ion{Zr}{vii}, the configurations included in the HFR model were 
4s$^2$4p$^4$, 
4s$^2$4p$^3$$n$p ($n$ = 5 -- 6), 
4s$^2$4p$^3$$n$f ($n$ = 4 -- 6), 
4s4p$^4$$n$d ($n$ = 4 -- 6), 
4s4p$^4$$n$s ($n$ = 5 -- 6), 
4p$^5$$n$p ($n$ = 5 -- 6), 
4p$^5$$n$f ($n$ = 4 -- 6), 
4s$^2$4p$^2$4d$^2$, 
4s$^2$4p$^2$4d5s, and 
4s$^2$4p$^2$5s$^2$ for the even parity, and 
4s4p$^5$, 
4s$^2$4p$^3$$n$d ($n$ = 4 -- 6), 
4s$^2$4p$^3$$n$s ($n$ = 5 -- 6), 
4s$^2$4p$^3$$n$g ($n$ = 5 -- 6), 
4s4p$^4$$n$p ($n$ = 5 -- 6), 
4s4p$^4$$n$f ($n$ = 4 -- 6), 
4p$^5$$n$s ($n$ = 5 -- 6), 
4p$^5$$n$d ($n$ = 4 -- 6), 
4s$^2$4p$^2$4d5p, and 
4s$^2$4p$^2$4d4f for the odd parity. The same core-polarization parameters as those used in \ion{Zr}{v} and \ion{Zr}{vi} calculations were 
considered while the radial integrals of 
4p$^4$, 
4p$^3$5p, 
4s4p$^5$, 
4p$^3$4d, and 
4p$^3$5s were optimized with the experimental energy levels taken from \citet{readeracquista1976,rahimullahetal1978, 
khanetal1983}. Although having established level values, the 
4p$^3$4f 
configuration was not fitted because it appeared 
very strongly mixed with experimentally unknown configurations such as 
4s4p$^4$4d, and 
4s$^2$4p$^2$4d$^2$ 
according to our HFR 
calculations. This semi-empirical process led to mean deviations of 695\,cm$^{-1}$ and 479\,cm$^{-1}$ for even and odd parities, 
respectively.

The parameters adopted in our computations are summarized in 
Tables \ref{tab:zriv:para} - \ref{tab:zrvii:para} while
computed and available experimental energies are compared in 
Tables \ref{tab:zriv:ener} - \ref{tab:zrvii:ener}, for \ion{Zr}{iv-vii}, respectively.
Tables \ref{tab:zriv:loggf} - \ref{tab:zrvii:loggf} 
give the HFR weighted oscillator 
strengths ($\log gf$) and transition probabilities ($gA$, in s$^{-1}$) together with the numerical values (in cm$^{-1}$) of the 
lower and upper energy levels and the corresponding wavelengths (in \AA). In the last column of each table, we also give the 
cancellation factor, $CF$, as defined by \citet{cowan1981}. We note that very low values of this factor (typically $<$ 0.05)
indicate 
strong cancellation effects in the calculation of line strengths. In these cases, the corresponding $gf$ and $gA$ values could 
be very inaccurate and therefore need to be considered with some care. However, very few of the transitions appearing in 
Tables \ref{tab:zriv:loggf} - \ref{tab:zrvii:loggf} are affected. 
These tables are provided via the registered GAVO T\"ubingen Oscillator Strengths Service 
(TOSS\footnote{\url{http://dc.g-vo.org/TOSS}}).

\subsection{Zr line identification and abundance analysis}
\label{sect:zrabundance}

\begin{figure*}
   \resizebox{\hsize}{!}{\includegraphics{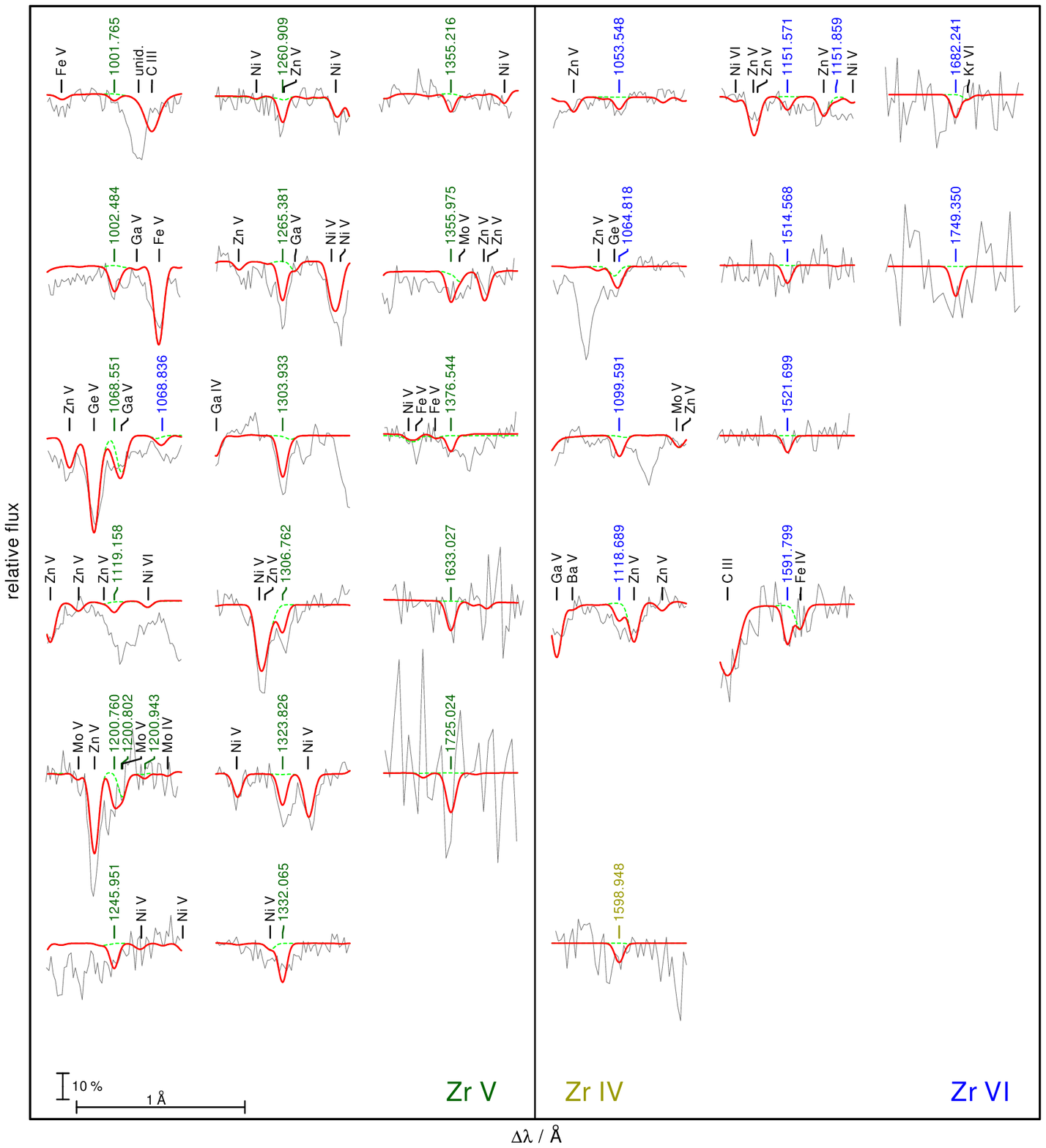}}
    \caption{Identified 
             \ion{Zr}{iv} (bottom of right panel), 
             \ion{Zr}{v} (left panel), and
             \ion{Zr}{vi} (right panel)  lines in the FUSE ($\lambda < 1188\,\mathrm{\AA}$)
             and HST/STIS observations of \re.
             The model (thick, red line) was calculated with an abundance of $\log \mathrm{Zr} = -3.5$.
             The dashed green spectrum was calculated without Zr.
             Prominent lines are marked, the Zr lines with their wavelengths from
             Tables \ref{tab:zriv:loggf} - \ref{tab:zrvi:loggf}.
            }
   \label{fig:zr}
\end{figure*}

In the FUSE and HST/STIS observations of \re, we identified 
\ion{Zr}{iv-vi} lines (Table\,\ref{tab:zrlineids}). The observation is well reproduced by our
model calculated with a mass fraction of $\log \mathrm{Zr} = -3.5 \pm 0.2$ (Fig.\,\ref{fig:zr}).
The \ion{Zr}{iv/v/vi} ionization equilibria are matched by our model.

\begin{table}\centering
\caption{Identified Zr lines in the UV spectrum of \re. 
         The wavelengths correspond to those in Tables \ref{tab:zriv:loggf} - \ref{tab:zrvi:loggf}.}
\label{tab:zrlineids}
\begin{tabular}{rlr@{.}ll}
\hline\hline
\noalign{\smallskip}
\multicolumn{2}{c}{} & \multicolumn{2}{c}{wavelength / \AA} & comment   \\
\hline
Zr & {\sc iv}        & 1598&948                             &           \\
Zr & {\sc v}         & 1001&765                             &           \\
\multicolumn{2}{c}{} & 1002&484                             &           \\
\multicolumn{2}{c}{} & 1068&551                             & blend \ion{Ga}{v} \\
\multicolumn{2}{c}{} & 1119&158                             & uncertain \\
\multicolumn{2}{c}{} & 1200&760                             &           \\
\multicolumn{2}{c}{} & 1245&951                             &           \\
\multicolumn{2}{c}{} & 1260&909                             &           \\
\multicolumn{2}{c}{} & 1265&381                             &           \\
\multicolumn{2}{c}{} & 1303&933                             &           \\
\multicolumn{2}{c}{} & 1306&762                             &           \\
\multicolumn{2}{c}{} & 1323&826                             &           \\
\multicolumn{2}{c}{} & 1332&065                             &           \\
\multicolumn{2}{c}{} & 1355&216                             &           \\
\multicolumn{2}{c}{} & 1355&975                             &           \\
\multicolumn{2}{c}{} & 1376&544                             &           \\
\multicolumn{2}{c}{} & 1633&027                             &           \\
\multicolumn{2}{c}{} & 1725&024                             & uncertain \\
Zr & {\sc vi}        & 1053&548                             &           \\
\multicolumn{2}{c}{} & 1064&818                             &           \\
\multicolumn{2}{c}{} & 1068&663                             & uncertain \\
\multicolumn{2}{c}{} & 1099&591                             &           \\
\multicolumn{2}{c}{} & 1118&689                             &           \\
\multicolumn{2}{c}{} & 1151&571                             &           \\
\multicolumn{2}{c}{} & 1514&568                             &           \\
\multicolumn{2}{c}{} & 1521&699                             &           \\
\multicolumn{2}{c}{} & 1591&799                             &           \\
\multicolumn{2}{c}{} & 1682&241                             &           \\
\multicolumn{2}{c}{} & 1749&350                             & uncertain \\
\hline
\end{tabular}
\end{table}

\begin{figure}
   \resizebox{\hsize}{!}{\includegraphics{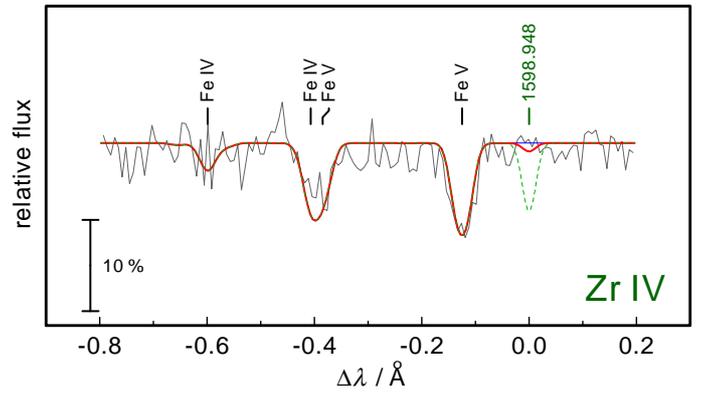}}
    \caption{Section of the STIS spectrum of \gb around \Jonw{Zr}{iv}{1598.948} compared with three 
     synthetic spectra 
     (thin, blue: no Zr, 
      thick, red: Zr mass fraction = $2.6\times 10^{-6}$,
      dashed green: Zr = $2.6\times 10^{-5}$).
           }
   \label{fig:zrg}
\end{figure}

In our synthetic spectra for \gb, \Ionw{Zr}{4}{1598.948} is the strongest line.
A comparison with the STIS spectrum shows that a Zr mass fraction of
$2.6\times 10^{-6}$ \citep[approximately 100 times solar, ][]{grevesseetal2015} is the upper detection limit
(Fig.\,\ref{fig:zrg}).

\section{Xenon}
\label{sect:xe}

\subsection{Oscillator-strength calculations for Xe {\small IV}, {\small V}, and {\small VII} ions}
\label{sect:xetrans}

New calculations of oscillator strengths and radiative transition probabilities in xenon ions were also performed using the 
HFR+CPOL method \citep{cowan1981,quinetetal1999,quinetetal2002}.

For \ion{Xe}{iv}, the multiconfiguration expansion included 
5s$^2$5p$^3$, 
5s$^2$5p$^2$6p, 
5s$^2$5p$^2$$n$f ($n$ = 4 -- 6), 
5s$^2$5p5d6s, 
5s$^2$5p5d6d, 
5s$^2$5p6s$^2$, 
5s$^2$5p5d$^2$, 
5s$^2$5p4f$^2$, 
5s5p$^3$6s, 
5s5p$^3$$n$d ($n$ = 5 -- 6), 
5s5p$^2$4f5d, and 
5p$^5$ for the odd parity, and 
5s5p$^4$, 
5s$^2$5p$^2$$n$d ($n$ = 5 -- 6), 
5s$^2$5p$^2$6s, 
5s$^2$5p$^2$$n$g ($n$ = 5 -- 6), 
5s$^2$5p5d6p, 
5s$^2$5p5d$n$f ($n$ = 4 -- 6), 
5s5p$^3$6p, 
5s5p$^3$$n$f ($n$ = 4 -- 6), and 
5s5p$^2$5d$^2$ for the even parity. 
The core-polarization effects were estimated with $\alpha$$_d$ = 0.88\,a.u\@. and $r_c$ = 0.86\,a.u\@. which correspond to a Pd-like 
\ion{Xe}{ix} ionic core. The former value was taken from \citet{fragaetal1976} while the latter one corresponds to the HFR mean value 
$<$$r$$>$ of the outermost core orbital (4d). The experimental energy levels published by \citet{saloman2004} were then used to optimize 
the radial parameters belonging to the 
5p$^3$, 
5p$^2$6p, 
5p$^2$4f, 
5s5p$^4$, 
5p$^2$5d, and 
5p$^2$6s configurations allowing us to reach average deviations between calculated and observed energies of 137cm$^{-1}$ and 
251\,cm$^{-1}$, for odd and even parities, respectively.

In the case of \ion{Xe}{v}, the following sets of configurations were considered in the HFR model: 
5s$^2$5p$^2$, 
5s$^2$5p6p, 
5s$^2$5p$n$f ($n$ = 4 -- 6), 
5s$^2$5d6s, 
5s$^2$5d6d, 
5s$^2$6s$^2$, 
5s$^2$5d$^2$, 
5s$^2$4f$^2$, 
5s$^2$5f$^2$, 
5s5p$^2$6s, 
5s5p$^2$$n$d ($n$ = 5 -- 6), 
5s5p6s6p, 
5s5p6p$n$d ($n$ = 5 -- 6), 
5s5p4f$n$d ($n$ = 5 -- 6), 
5p$^4$, 
5p$^3$6p, and
5p$^3$$n$f ($n$ = 4 -- 6) 
for the even parity, and 
5s5p$^3$, 5s$^2$5p$n$d ($n$ = 5 -- 6), 
5s$^2$5p$n$s ($n$ = 6 -- 7), 
5s$^2$5p$n$g ($n$ = 5 -- 6), 
5s$^2$5d6p, 
5s$^2$5d$n$f ($n$ = 4 -- 6), 
5s5p$^2$6p, 
5s5p$^2$$n$f ($n$ = 4 -- 6), 
5s5p6s$n$d ($n$ = 5 -- 6), 
5s5p5d6d, 
5s5p6s$^2$, 
5s5p5d$^2$, 
5p$^3$6s, and
5p$^3$$n$d ($n$ = 5 -- 6) 
for the odd parity. The same core-polarization parameters as those used for \ion{Xe}{iv} were used and the 
experimental energy levels reported by \citet{saloman2004} and \citet{rainerietal2009} were incorporated into 
the semi-empirical fit to adjust the radial integrals corresponding to the 
5p$^2$, 
5p6p, 
5p4f, 
5s5p$^3$, 
5p5d, 
5p6d, 
5p6s, and 
5p7s configurations. In this process, we found mean deviations equal to 144\,cm$^{-1}$ in the even parity 
and 110\,cm$^{-1}$ in the odd parity.

For \ion{Xe}{vi}, we used the same atomic data as those considered in one of our previous papers \citep{rauchetal2015xe}. 
More precisely, the radiative rates were taken from the work of \citet{gallardoetal2015} who performed HFR+CPOL 
calculations including 35 odd-parity and 34 even-parity configurations, that is, 
5s$^2$$n$p ($n$ = 5 -- 8), 
5s$^2$$n$f ($n$ = 4 -- 8), 
5s$^2$$n$h ($n$ = 6 -- 8), 
5s$^2$8k, 5p$^2$$n$p ($n$ = 6 -- 8), 
5p$^2$$n$f ($n$ = 4 -- 8), 
5p$^2$$n$h ($n$ = 6 -- 8), 
5p$^2$8k, 
5s5p6s, 
5s5p$n$d ($n$ = 5 -- 6), 
5s5p$n$g ($n$ = 5 -- 6), 
5p$^3$, 
5s5d$n$f ($n$ = 4 -- 5), 
5s6s$n$f ($n$ = 4 -- 5), and   
5s5p$^2$, 
5s$^2$$n$s ($n$ = 6 -- 8), 
5s$^2$$n$d ($n$ = 5 -- 8), 
5s$^2$$n$g ($n$ = 5 -- 8), 
5s$^2$$n$i ($n$ = 7 -- 8), 
5p$^2$$n$d ($n$ = 5 -- 8), 
5p$^2$$n$s ($n$ = 6 -- 8), 
5p$^2$$n$g ($n$ = 5 -- 8), 
5p$^2$$n$i ($n$ = 7 -- 8), 
5s5p$n$f ($n$ = 4 -- 6), 
5s4f$^2$, 
5s5f$^2$, 
5s5p6p, 
4d$^9$5p$^4$, respectively. In this latter study, the core-polarization effects were considered with two different ionic cores, 
that is, a Cd-like \ion{Xe}{vii} core with $\alpha_\mathrm{d}$ = 5.80\,a.u. for the 
5s$^2$nl -- 5s$^2$n'l' 
transitions, and a Pd-like 
\ion{Xe}{ix} core with $\alpha_\mathrm{d}$ = 0.99\,a.u\@. for all the other transitions. 
In their semi-empirical least-squares fitting process, 
\citet{gallardoetal2015} achieved standard deviations with experimental energy levels of 149\,cm$^{-1}$ in the odd parity and 
154\,cm$^{-1}$ in the even parity.

Finally, for \ion{Xe}{vii}, we used the same model as the one considered by \citet{biemontetal2007} extending the set of 
oscillator strengths to weaker transitions (up to log $gf$ $>-8$). As a reminder, these authors explicitly retained the following configurations in 
their configuration interaction expansions:
5s$^2$, 
5p$^2$, 
5d$^2$, 
4f$^2$, 
4f$n$p ($n$ = 5 -- 6), 
4f6f, 
4f6h, 
5s6s, 
5s$n$d ($n$ = 5 -- 6), 
5s$n$g ($n$ = 5 -- 6), 
5p$n$f ($n$ = 5 -- 6), 
5p6p, 
5p6h, 
5d6s, 
5d6d, and
5d$n$g ($n$ = 5 -- 6) for the even parity, and 
5s$n$p ($n$ = 5 -- 6), 
5s$n$f ($n$ = 4 -- 6), 
5s6h, 
4f6s, 
4f$n$d ($n$ = 5 -- 6), 
4f$n$g ($n$ = 5 -- 6), 
5p6s, 
5p$n$d ($n$ = 5 -- 6), 
5p$n$g ($n$ = 5 -- 6), 
5d6p, and
5d$n$f ($n$ = 5 -- 6), 5d6h for the odd parity. 
The same ionic core parameters as those used for Xe IV and Xe V ions were considered and all the experimental energy 
levels published by \citet{saloman2004} were included in the semi-empirical optimization of the radial parameters belonging 
to the 
5s$^2$, 
5s6s, 
5s5d, 
5s6d, 
5p$^2$, 
4f5p, 
5s5p, 
5s6p, 
5s4f, 
5s5f, 
5p6s, and 
5p5d configurations giving rise to standard deviations of 377\,cm$^{-1}$ and 250\,cm$^{-1}$ for even- and odd-parity levels, 
respectively.

The radial parameters used in our computations are summarized in 
Tables \ref{tab:xeiv:para} - \ref{tab:xev:para}
for the \ion{Xe}{iv-v} ions, respectively. 
The calculated energy levels are compared with available experimental values in 
Tables \ref{tab:xeiv:ener} - \ref{tab:xev:ener}
while the HFR weighted 
oscillator strengths (log $gf$) and transition probabilities ($gA$ in s$^{-1}$) are reported in 
Tables \ref{tab:xeiv:loggf} - \ref{tab:xevii:loggf} for the \ion{Xe}{iv-v} and  \ion{}{vii} ions, respectively. 
In the latter tables, we also give the numerical values (in cm$^{-1}$) of lower and upper energy levels 
of each transition together with the corresponding wavelength (in \AA) and the 
$CF$, as introduced in Sect.\,\ref{sect:zrtrans}. These tables are provided via TOSS.

\subsection{Xe line identification and abundance analysis}
\label{sect:xeabundance}

\begin{figure*}
   \resizebox{\hsize}{!}{\includegraphics{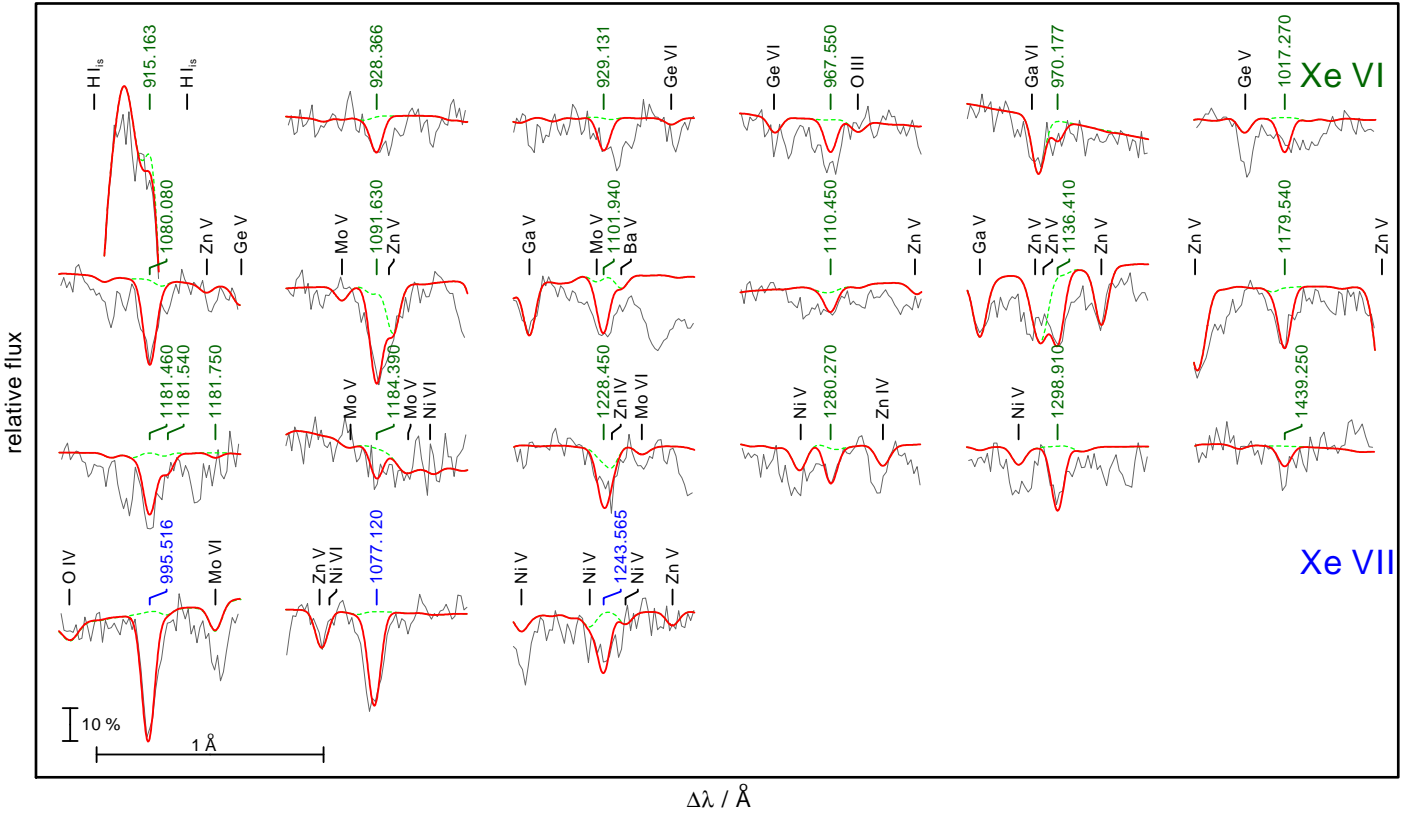}}
    \caption{Identified 
             \ion{Xe}{vi} (top three rows) and
             \ion{Xe}{vii} (bottom row)  lines in the FUSE ($\lambda < 1188\,\mathrm{\AA}$)
             and HST/STIS observations of \re.
             The model (thick, red line) was calculated with an abundance of $\log \mathrm{Xe} = -3.9$.
             The dashed, green spectrum was calculated without Xe.
             Prominent lines are marked (`is' denotes interstellar origin), and 
             the Xe lines are labelled with their wavelengths given by
             \citet{gallardoetal2015} and in Table \ref{tab:xevii:loggf}.
            }
   \label{fig:xe}
\end{figure*}

In the FUSE and HST/STIS observations of \re, we identified 
\ion{Xe}{vi-vii} lines (Table\,\ref{tab:xelineids}). The observation is well reproduced by our
model, calculated with a mass fraction of $\log \mathrm{Xe} = -3.9 \pm 0.2$ (Fig.\,\ref{fig:xe}).
This is a factor of two higher than that previously determined by \citet[][$\log \mathrm{Xe} = -4.2 \pm 0.6$]{werneretal2012}
but agrees within their given error limits.
The \ion{Xe}{vi/vii} ionization equilibrium is matched by our model.

\begin{table}\centering
\caption{Identified Xe lines in the UV spectrum of \re. 
         The wavelengths correspond to those given in \citet{gallardoetal2015} and in Table \ref{tab:xevii:loggf}
         for \ion{Xe}{vi} and \ion{Xe}{vii}, respectively.}
\label{tab:xelineids}
\begin{tabular}{rlr@{.}ll}
\hline\hline
\noalign{\smallskip}
\multicolumn{2}{c}{} & \multicolumn{2}{c}{wavelength / \AA} & comment   \\
\hline
Xe & {\sc vi}        &  915&163                                              & weak      \\
\multicolumn{2}{c}{} &  928&366\tablefootmark{a}                             &           \\
\multicolumn{2}{c}{} &  929&131\tablefootmark{b}                             &           \\
\multicolumn{2}{c}{} &  967&550\tablefootmark{a}                             &           \\
\multicolumn{2}{c}{} &  970&177                                              & weak      \\
\multicolumn{2}{c}{} & 1017&270\tablefootmark{b}                             &           \\
\multicolumn{2}{c}{} & 1080&080\tablefootmark{a}                             &           \\
\multicolumn{2}{c}{} & 1091&630\tablefootmark{a}                             &           \\
\multicolumn{2}{c}{} & 1101&940\tablefootmark{a}                             &           \\
\multicolumn{2}{c}{} & 1110&450                                              & weak      \\
\multicolumn{2}{c}{} & 1136&410\tablefootmark{a}                             &           \\
\multicolumn{2}{c}{} & 1179&540\tablefootmark{a}                             &           \\
\multicolumn{2}{c}{} & 1181&390\tablefootmark{a}                             &           \\
\multicolumn{2}{c}{} & 1181&540                                              & blend with \Jonw{Xe}{vi}{1181.390}          \\
\multicolumn{2}{c}{} & 1184&390\tablefootmark{a}                             & uncertain \\
\multicolumn{2}{c}{} & 1228&450                                              &           \\
\multicolumn{2}{c}{} & 1280&270                                              &           \\
\multicolumn{2}{c}{} & 1298&910\tablefootmark{b}                             &           \\
\multicolumn{2}{c}{} & 1439&250                                              &           \\
Xe & {\sc vii}       &  995&516\tablefootmark{a}                             &           \\
\multicolumn{2}{c}{} & 1077&120\tablefootmark{a}                             &           \\
\multicolumn{2}{c}{} & 1243&565                                              &           \\
\hline
\end{tabular}
\tablefoot{
\tablefoottext{a}{identified by \citet{werneretal2012}},
\tablefoottext{b}{identified by \citet{rauchetal2015xe}}
}
\end{table}

\section{Results and conclusions}
\label{sect:results}

To search for Al lines in the observed UV spectrum of \re, we created an extended Al model atom for our 
NLTE model-atmosphere calculations. We could only identify \Ionww{Al}{3}{1384.130} (Sect.\,\ref{sect:al}),
that was well suited to measure the Al abundance. 
It is reproduced at a solar value ($-4.28 \pm 0.2$, mass fraction).
This needs to be verified once better observations are available.

\begin{figure}
   \resizebox{\hsize}{!}{\includegraphics{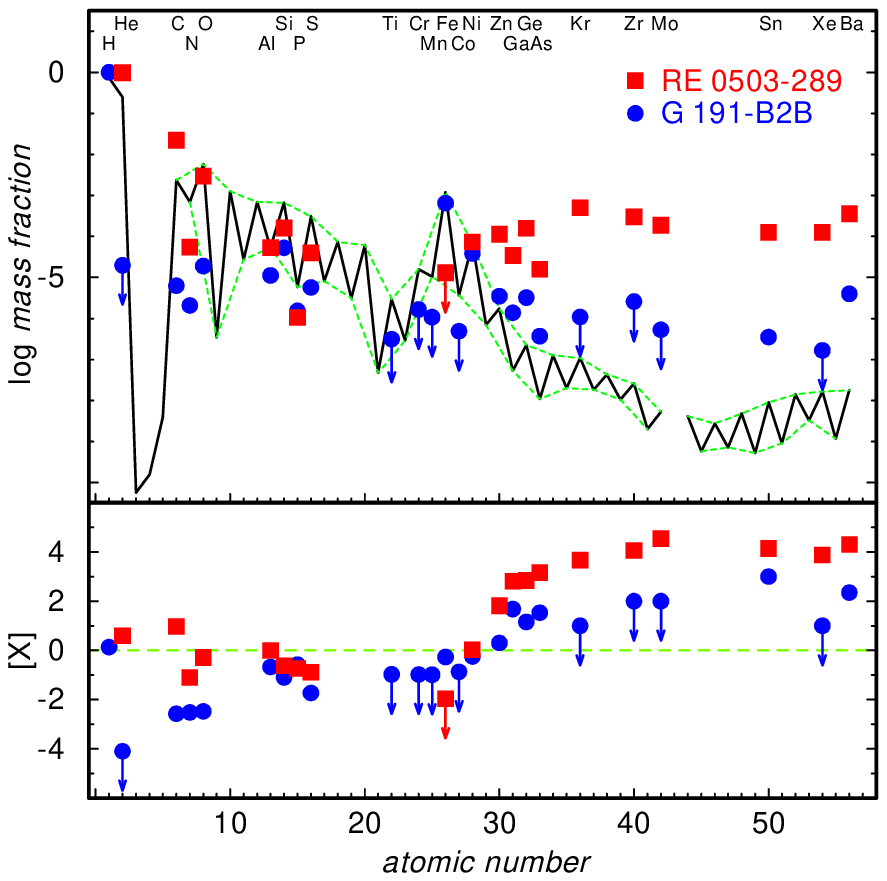}}
    \caption{Solar abundances \citep[thick line; the dashed lines
             connect the elements with even and with odd atomic number]{asplundetal2009,scottetal2015a,scottetal2015b,grevesseetal2015}
             compared with the determined photospheric abundances of 
             \gb \citep[blue circles,][]{rauchetal2013} and
             \re \citep[red squares,][and this work]{dreizlerwerner1996,rauchetal2012ge, rauchetal2014zn, rauchetal2014ba, rauchetal2015xe, rauchetal2015ga, rauchetal2016mo, rauchetal2016kr}.
             The uncertainties of the WD abundances are, 
in general, approximately 0.2\,dex. The arrows indicate upper limits.
             Top panel: Abundances given as logarithmic mass fractions.
             Bottom panel: Abundance ratios to respective solar values, 
                           [X] denotes log (fraction\,/\,solar fraction) of species X.
                           The dashed green line indicates solar abundances.
            }
   \label{fig:X}
\end{figure}

We identified \ion{Zr}{iv - vi} lines in the observed high-resolution UV spectra \re (Table\,\ref{tab:zrlineids}).
These were well modeled using
our newly calculated \ion{Zr}{iv-vii} oscillator strengths. We determined a photospheric abundance of 
$\log\,\mathrm{Zr} = -3.52 \pm 0.2$ (mass fraction, $1.5 - 4.8\,\times\,10^{-4}$,  5775 --  14\,480 times the solar abundance).
This highly supersolar Zr abundance corresponds to the high abundances of other trans-iron elements in \re (Fig.\,\ref{fig:X}).
The \ion{Zr}{iv/v/vi} ionization equilibria are well matched by our model (\Teffw{70\,000}, \loggw{7.5}).

In addition to the previously discovered \ion{Xe}{vi - vii} lines in the UV spectrum of \re, we identified
five \ion{new Xe}{vi} lines. All identified Xe lines are well matched by our model with an abundance of
$\log\,\mathrm{Xe} = -3.88 \pm 0.2$ (mass fraction, $0.8 - 2.1\,\times\,10^{-4}$,  4985 --  12\,520 times the solar abundance).
This highly supersolar Xe abundance is in line with abundances of other trans-iron elements in \re (Fig.\,\ref{fig:X}).

\begin{figure*}
   \resizebox{\hsize}{!}{\includegraphics{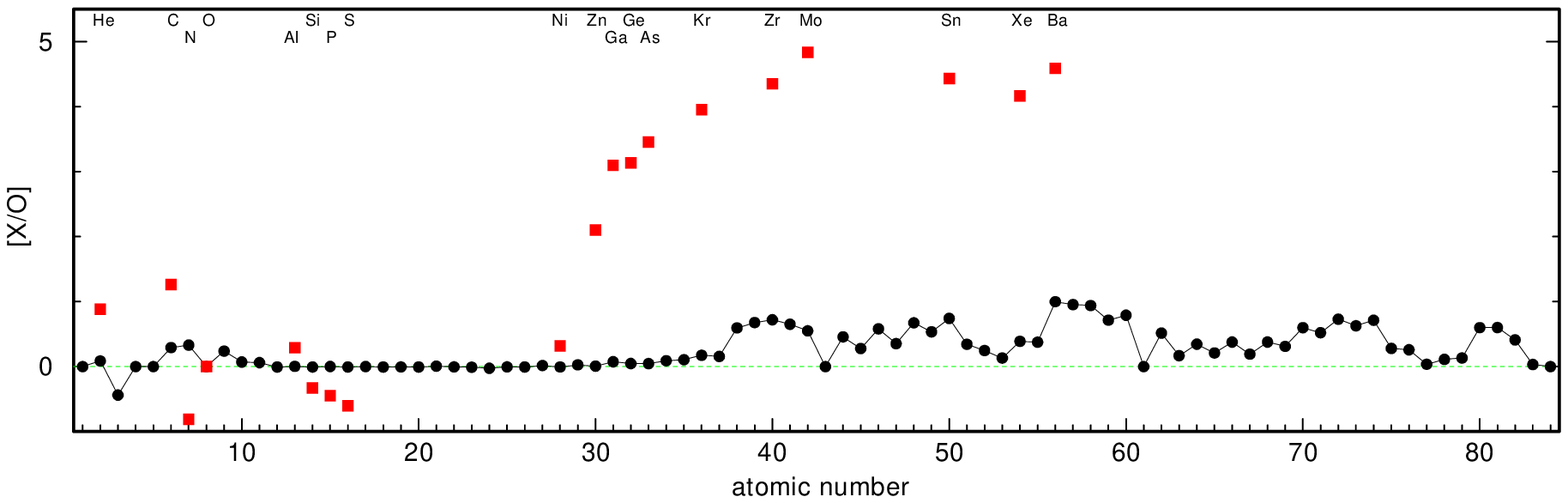}}
    \caption{Determined photospheric abundances of \re (cf\@. Fig\,\ref{fig:X}) compared with predictions for surface abundances of 
             \citet[][for an asymptotic giant branch (AGB) star with
             $M_\mathrm{initial} = 1.5\,M_\odot$, $M_\mathrm{final} = 0.585\,M_\odot$, metallicity $Z = 0.014$]{karakaslugaro2016}.
             [X/O] denotes the normalized $\log \left[\mathrm{(fraction\,of\,X\,/\,solar\,fraction\,of\,X) / 
                                                              (fraction\,of\,O\,/\,solar\,fraction\,of\,O)}\right]$ mass ratio.
             The dashed green line indicates the solar ratio.
            }
   \label{fig:klxo}
\end{figure*}

The amount of trans-iron elements in the photosphere of \re strongly exceeds the yields of nucleosynthesis
on the asymptotic giant branch (Fig\,\ref{fig:klxo}). It is likely that radiative levitation is working efficiently in \re
\citep{rauchetal2016mo}, increasing abundances by up to 4\,dex compared with solar values.

The identification of lines of Zr and Xe and their precise abundance determinations only  became possible 
after reliable transition probabilities for \ion{Zr}{iv - vii}, \ion{Xe}{iv-v}, and \ion{Xe}{vii}
were computed. Calculations for other, highly-ionized trans-iron elements are necessary to search for 
their lines and to measure their abundances.

The search for Zr and Xe lines in the UV spectrum of \gb was entirely negative. We established an upper
Zr abundance limit of approximately 100 times solar and confirmed the previously found upper limit for
Xe of approximately 10 times solar \citep{rauchetal2016mo}.

\begin{acknowledgements}
TR and DH are supported by the German Aerospace Center (DLR, grants 05\,OR\,1507 and 50\,OR\,1501, respectively).
The GAVO project had been supported by the Federal Ministry of Education and
Research (BMBF) 
at T\"ubingen (05\,AC\,6\,VTB, 05\,AC\,11\,VTB) and is funded
at Heidelberg (05\,AC\,11\,VH3).
Financial support from the Belgian FRS-FNRS is also acknowledged. 
PQ is research director of this organization.
Some of the data presented in this paper were obtained from the
Mikulski Archive for Space Telescopes (MAST). STScI is operated by the
Association of Universities for Research in Astronomy, Inc., under NASA
contract NAS5-26555. Support for MAST for non-HST data is provided by
the NASA Office of Space Science via grant NNX09AF08G and by other
grants and contracts. 
This research has made use of 
NASA's Astrophysics Data System and
the SIMBAD database, operated at CDS, Strasbourg, France.
The TOSS service (\url{http://dc.g-vo.org/TOSS}) that provides weighted oscillator strengths and transition probabilities
was constructed as part of the activities of the German Astrophysical Virtual Observatory.
\end{acknowledgements}

\bibliographystyle{aa}
\bibliography{29794}

\begin{thebibliography}{43}
\expandafter\ifx\csname natexlab\endcsname\relax\def\natexlab#1{#1}\fi

\bibitem[{{Asplund} {et~al.}(2009){Asplund}, {Grevesse}, {Sauval}, \&
  {Scott}}]{asplundetal2009}
{Asplund}, M., {Grevesse}, N., {Sauval}, A.~J., \& {Scott}, P. 2009, \araa, 47,
  481

\bibitem[{{Bi{\'e}mont} {et~al.}(2007){Bi{\'e}mont}, {Clar}, {Fivet}, {Garnir},
  {Palmeri}, {Quinet}, \& {Rostohar}}]{biemontetal2007}
{Bi{\'e}mont}, {\'E}., {Clar}, M., {Fivet}, V., {et~al.} 2007, European
  Physical Journal D, 44, 23

\bibitem[{{Bohlin}(2007)}]{bohlin2007}
{Bohlin}, R.~C. 2007, in Astronomical Society of the Pacific Conference Series,
  Vol. 364, The Future of Photometric, Spectrophotometric and Polarimetric
  Standardization, ed. C.~{Sterken}, 315

\bibitem[{{Cowan}(1981)}]{cowan1981}
{Cowan}, R.~D. 1981, {The theory of atomic structure and spectra} (Berkeley,
  CA, University of California Press)

\bibitem[{{Cowley}(1970)}]{cowley1970}
{Cowley}, C.~R. 1970, {The theory of stellar spectra} (Gordon \& Breach, New
  York)

\bibitem[{{Cowley}(1971)}]{cowley1971}
{Cowley}, C.~R. 1971, The Observatory, 91, 139

\bibitem[{{Dreizler} \& {Werner}(1996)}]{dreizlerwerner1996}
{Dreizler}, S. \& {Werner}, K. 1996, \aap, 314, 217

\bibitem[{{Fraga} {et~al.}(1976){Fraga}, {Karwowski}, \&
  {Saxena}}]{fragaetal1976}
{Fraga}, S., {Karwowski}, J., \& {Saxena}, K.~M.~S. 1976, {Handbook of Atomic
  Data} (Elsevier, Amsterdam)

\bibitem[{{Gallardo} {et~al.}(2015){Gallardo}, {Raineri}, {Reyna Almandos},
  {Pagan}, \& {Abrah{\~a}o}}]{gallardoetal2015}
{Gallardo}, M., {Raineri}, M., {Reyna Almandos}, J., {Pagan}, C.~J.~B., \&
  {Abrah{\~a}o}, R.~A. 2015, \apjs, 216, 11

\bibitem[{{Grevesse} {et~al.}(2015){Grevesse}, {Scott}, {Asplund}, \&
  {Sauval}}]{grevesseetal2015}
{Grevesse}, N., {Scott}, P., {Asplund}, M., \& {Sauval}, A.~J. 2015, \aap, 573,
  A27

\bibitem[{{Holberg} {et~al.}(1998){Holberg}, {Barstow}, \&
  {Sion}}]{holbergetal1998}
{Holberg}, J.~B., {Barstow}, M.~A., \& {Sion}, E.~M. 1998, \apjs, 119, 207

\bibitem[{{Hubeny} {et~al.}(1994){Hubeny}, {Hummer}, \&
  {Lanz}}]{hubenyetal1994}
{Hubeny}, I., {Hummer}, D.~G., \& {Lanz}, T. 1994, \aap, 282, 151

\bibitem[{{Hummer} \& {Mihalas}(1988)}]{hummermihalas1988}
{Hummer}, D.~G. \& {Mihalas}, D. 1988, \apj, 331, 794

\bibitem[{{Karakas} \& {Lugaro}(2016)}]{karakaslugaro2016}
{Karakas}, A.~I. \& {Lugaro}, M. 2016, \apj, 825, 26

\bibitem[{{Khan} {et~al.}(1983){Khan}, {Chaghtai}, \&
  {Rahimullah}}]{khanetal1983}
{Khan}, Z.~A., {Chaghtai}, M.~S.~Z., \& {Rahimullah}, K. 1983, Journal of
  Physics B: Atomic and Molecular Physics, 16, 1685

\bibitem[{{Khan} {et~al.}(1981){Khan}, {Rahimullah}, \&
  {Chaghtai}}]{khanetal1981}
{Khan}, Z.~A., {Rahimullah}, K., \& {Chaghtai}, M.~S.~Z. 1981, Physica Scripta,
  23, 843

\bibitem[{{Lemoine} {et~al.}(2002){Lemoine}, {Vidal-Madjar}, {H{\'e}brard},
  {D{\'e}sert}, {Ferlet}, {Lecavelier des {\'E}tangs}, {Howk}, {Andr{\'e}},
  {Blair}, {Friedman}, {Kruk}, {Lacour}, {Moos}, {Sembach}, {Chayer},
  {Jenkins}, {Koester}, {Linsky}, {Wood}, {Oegerle}, {Sonneborn}, \&
  {York}}]{lemoineetal2002}
{Lemoine}, M., {Vidal-Madjar}, A., {H{\'e}brard}, G., {et~al.} 2002, \apjs,
  140, 67

\bibitem[{{McCook} \& {Sion}(1999{\natexlab{a}})}]{mccooksion1999}
{McCook}, G.~P. \& {Sion}, E.~M. 1999{\natexlab{a}}, \apjs, 121, 1

\bibitem[{{McCook} \& {Sion}(1999{\natexlab{b}})}]{mccooksion1999cat}
{McCook}, G.~P. \& {Sion}, E.~M. 1999{\natexlab{b}}, VizieR Online Data
  Catalog, 3210, 0

\bibitem[{{M\"uller-Ringat}(2013)}]{muellerringatPhD2013}
{M\"uller-Ringat}, E. 2013, Dissertation, University of T\"ubingen, Germany,
  \url{http://www.ivoa.net/documents/SimDM/index.html}

\bibitem[{{Quinet} {et~al.}(2002){Quinet}, {Palmeri}, {Bi{\'e}mont}, {Li},
  {Zhang}, \& {Svanberg}}]{quinetetal2002}
{Quinet}, P., {Palmeri}, P., {Bi{\'e}mont}, {\'E}., {et~al.} 2002, J. Alloys
  Comp., 344, 255

\bibitem[{{Quinet} {et~al.}(1999){Quinet}, {Palmeri}, {Bi{\'e}mont}, {McCurdy},
  {Rieger}, {Pinnington}, {Wickliffe}, \& {Lawler}}]{quinetetal1999}
{Quinet}, P., {Palmeri}, P., {Bi{\'e}mont}, {\'E}., {et~al.} 1999, \mnras, 307,
  934

\bibitem[{{Rahimullah} {et~al.}(1978){Rahimullah}, {Chaghtai}, \&
  {Khatoon}}]{rahimullahetal1978}
{Rahimullah}, K., {Chaghtai}, M.~S.~Z., \& {Khatoon}, S. 1978, Physica Scripta,
  18, 96

\bibitem[{{Raineri} {et~al.}(2009){Raineri}, {Gallardo}, {Padilla}, \& {Reyna
  Almandos}}]{rainerietal2009}
{Raineri}, M., {Gallardo}, M., {Padilla}, S., \& {Reyna Almandos}, J. 2009,
  Journal of Physics B Atomic Molecular Physics, 42, 205004

\bibitem[{{Rauch} \& {Deetjen}(2003)}]{rauchdeetjen2003}
{Rauch}, T. \& {Deetjen}, J.~L. 2003, in Astronomical Society of the Pacific
  Conference Series, Vol. 288, Stellar Atmosphere Modeling, ed. I.~{Hubeny},
  D.~{Mihalas}, \& K.~{Werner}, 103

\bibitem[{{Rauch} {et~al.}(2015{\natexlab{a}}){Rauch}, {Hoyer}, {Quinet},
  {Gallardo}, \& {Raineri}}]{rauchetal2015xe}
{Rauch}, T., {Hoyer}, D., {Quinet}, P., {Gallardo}, M., \& {Raineri}, M.
  2015{\natexlab{a}}, \aap, 577, A88

\bibitem[{{Rauch} {et~al.}(2016{\natexlab{a}}){Rauch}, {Quinet}, {Hoyer},
  {Werner}, {Demleitner}, \& {Kruk}}]{rauchetal2016mo}
{Rauch}, T., {Quinet}, P., {Hoyer}, D., {et~al.} 2016{\natexlab{a}}, \aap, 587,
  A39

\bibitem[{{Rauch} {et~al.}(2016{\natexlab{b}}){Rauch}, {Quinet}, {Hoyer},
  {Werner}, {Richter}, {Kruk}, \& {Demleitner}}]{rauchetal2016kr}
{Rauch}, T., {Quinet}, P., {Hoyer}, D., {et~al.} 2016{\natexlab{b}}, \aap, 590,
  A128

\bibitem[{{Rauch} {et~al.}(2012){Rauch}, {Werner}, {Bi{\'e}mont}, {Quinet}, \&
  {Kruk}}]{rauchetal2012ge}
{Rauch}, T., {Werner}, K., {Bi{\'e}mont}, {\'E}., {Quinet}, P., \& {Kruk},
  J.~W. 2012, \aap, 546, A55

\bibitem[{{Rauch} {et~al.}(2013){Rauch}, {Werner}, {Bohlin}, \&
  {Kruk}}]{rauchetal2013}
{Rauch}, T., {Werner}, K., {Bohlin}, R., \& {Kruk}, J.~W. 2013, \aap, 560, A106

\bibitem[{{Rauch} {et~al.}(2014{\natexlab{a}}){Rauch}, {Werner}, {Quinet}, \&
  {Kruk}}]{rauchetal2014zn}
{Rauch}, T., {Werner}, K., {Quinet}, P., \& {Kruk}, J.~W. 2014{\natexlab{a}},
  \aap, 564, A41

\bibitem[{{Rauch} {et~al.}(2014{\natexlab{b}}){Rauch}, {Werner}, {Quinet}, \&
  {Kruk}}]{rauchetal2014ba}
{Rauch}, T., {Werner}, K., {Quinet}, P., \& {Kruk}, J.~W. 2014{\natexlab{b}},
  \aap, 566, A10

\bibitem[{{Rauch} {et~al.}(2015{\natexlab{b}}){Rauch}, {Werner}, {Quinet}, \&
  {Kruk}}]{rauchetal2015ga}
{Rauch}, T., {Werner}, K., {Quinet}, P., \& {Kruk}, J.~W. 2015{\natexlab{b}},
  \aap, 577, A6

\bibitem[{{Reader} \& {Acquista}(1976)}]{readeracquista1976}
{Reader}, J. \& {Acquista}, N. 1976, Journal of the Optical Society of America
  (1917-1983), 66, 896

\bibitem[{{Reader} \& {Acquista}(1979)}]{readeracquista1979}
{Reader}, J. \& {Acquista}, N. 1979, Journal of the Optical Society of America
  (1917-1983), 69, 239

\bibitem[{{Reader} \& {Acquista}(1997)}]{readeracquista1997}
{Reader}, J. \& {Acquista}, N. 1997, Journal of the Optical Society of America
  B Optical Physics, 14, 1328

\bibitem[{{Reader} \& {Lindsay}(2016)}]{readerlindsay2016}
{Reader}, J. \& {Lindsay}, M.~D. 2016, \physscr, 91, 025401

\bibitem[{{Saloman}(2004)}]{saloman2004}
{Saloman}, E.~B. 2004, Journal of Physical and Chemical Reference Data, 33, 765

\bibitem[{{Scott} {et~al.}(2015{\natexlab{a}}){Scott}, {Asplund}, {Grevesse},
  {Bergemann}, \& {Sauval}}]{scottetal2015b}
{Scott}, P., {Asplund}, M., {Grevesse}, N., {Bergemann}, M., \& {Sauval}, A.~J.
  2015{\natexlab{a}}, \aap, 573, A26

\bibitem[{{Scott} {et~al.}(2015{\natexlab{b}}){Scott}, {Grevesse}, {Asplund},
  {Sauval}, {Lind}, {Takeda}, {Collet}, {Trampedach}, \&
  {Hayek}}]{scottetal2015a}
{Scott}, P., {Grevesse}, N., {Asplund}, M., {et~al.} 2015{\natexlab{b}}, \aap,
  573, A25

\bibitem[{{Werner} {et~al.}(2003){Werner}, {Deetjen}, {Dreizler}, {Nagel},
  {Rauch}, \& {Schuh}}]{werneretal2003}
{Werner}, K., {Deetjen}, J.~L., {Dreizler}, S., {et~al.} 2003, in Astronomical
  Society of the Pacific Conference Series, Vol. 288, Stellar Atmosphere
  Modeling, ed. I.~{Hubeny}, D.~{Mihalas}, \& K.~{Werner}, 31

\bibitem[{{Werner} {et~al.}(2012{\natexlab{a}}){Werner}, {Dreizler}, \&
  {Rauch}}]{tmap2012}
{Werner}, K., {Dreizler}, S., \& {Rauch}, T. 2012{\natexlab{a}}, {TMAP:
  T{\"u}bingen NLTE Model-Atmosphere Package}, Astrophysics Source Code Library
  [record ascl:1212.015]

\bibitem[{{Werner} {et~al.}(2012{\natexlab{b}}){Werner}, {Rauch}, {Ringat}, \&
  {Kruk}}]{werneretal2012}
{Werner}, K., {Rauch}, T., {Ringat}, E., \& {Kruk}, J.~W. 2012{\natexlab{b}},
  \apjl, 753, L7

\end{thebibliography}

\begin{appendix}
\onecolumn

\section{Additional tables for zirconium}
\label{app:addtabzr}

\begin{longtable}{llrrrl}
\caption{\label{tab:zriv:para}Radial parameters (in cm$^{-1}$) adopted for the calculations in \ion{Zr}{iv}.} \\
\hline\hline
\noalign{\smallskip}
Configuration & Parameter & HFR & Fitted & Ratio & Note$^a$ \\ 
\noalign{\smallskip}
\hline
\endfirsthead
\caption{continued.}\\
\hline\hline
\noalign{\smallskip}
Configuration & Parameter & HFR & Fitted & Ratio & Note$^a$ \\
\noalign{\smallskip}
\hline
\noalign{\smallskip}
\endhead
\hline
\noalign{\smallskip}
\endfoot
\noalign{\smallskip}
\multicolumn{6}{c}{Even parity} \\
4d               & E$_{av}$          &    3588 &   3569 &       &    \\
                 & $\zeta$$_{4d}$    &     515 &    503 & 0.975 &    \\
5d               & E$_{av}$          &  149752 & 147800 &       &    \\
                 & $\zeta$$_{5d}$    &     116 &    136 & 1.172 &    \\
6d               & E$_{av}$          &  200505 & 198248 &       &    \\
                 & $\zeta$$_{6d}$    &      53 &     64 & 1.209 &    \\
5s               & E$_{av}$          &   42289 &  41703 &       &    \\
\hline
\end{longtable}
$^a$ F: Fixed parameter value.\\
\noindent

\begin{longtable}{llrrrl}
\caption{\label{tab:zrvv:para}Radial parameters (in cm$^{-1}$) adopted for the calculations in \ion{Zr}{v}.} \\
\hline\hline
\noalign{\smallskip}
Configuration & Parameter & HFR & Fitted & Ratio & Note$^a$ \\ 
\noalign{\smallskip}
\hline
\endfirsthead
\caption{continued.}\\
\hline\hline
\noalign{\smallskip}
Configuration & Parameter & HFR & Fitted & Ratio & Note$^a$ \\
\noalign{\smallskip}
\hline
\noalign{\smallskip}
\endhead
\hline
\noalign{\smallskip}
\endfoot
\noalign{\smallskip}
\multicolumn{6}{c}{Even parity} \\
4p$^6$           & E$_{av}$          &   17448 &  17850 &       &    \\
4p$^5$5p         & E$_{av}$          &  387307 & 386625 &       &    \\
                 & $\zeta$$_{4p}$    &    9939 &  10310 & 1.037 &    \\
                 & $\zeta$$_{5p}$    &    1865 &   2181 & 1.169 &    \\
                 & F$^2$(4p,5p)      &   22698 &  18416 & 0.811 &    \\    
                 & G$^0$(4p,5p)      &    4681 &   3978 & 0.850 & R1 \\
                 & G$^2$(4p,5p)      &    6345 &   5392 & 0.850 & R1 \\
\hline
\end{longtable}
$^a$ F: Fixed parameter value; Rn: ratios of these parameters have been fixed in the fitting process
\noindent

\begin{longtable}{llrrrl}
\caption{\label{tab:zrvi:para}Radial parameters (in cm$^{-1}$) adopted for the calculations in \ion{Zr}{vi}.} \\
\hline\hline
\noalign{\smallskip}
Configuration & Parameter & HFR & Fitted & Ratio & Note$^a$ \\ 
\noalign{\smallskip}
\hline
\endfirsthead
\caption{continued.}\\
\hline\hline
\noalign{\smallskip}
Configuration & Parameter & HFR & Fitted & Ratio & Note$^a$ \\
\noalign{\smallskip}
\hline
\noalign{\smallskip}
\endhead
\hline
\noalign{\smallskip}
\endfoot
\noalign{\smallskip}
\multicolumn{6}{c}{Odd parity} \\
4p$^5$           & E$_{av}$          &   22997 &  23322 &       &    \\
                 & $\zeta$$_{4p}$    &   10007 &  10580 & 1.057 &    \\
4p$^4$5p         & E$_{av}$          &  461912 & 446765 &       &    \\
                 & F$^2$(4p,4p)      &   84088 &  79559 & 0.946 &    \\
                 & $\alpha$          &       0 &   -651 &       &    \\
                 & $\zeta$$_{4p}$    &   10577 &  10907 & 1.031 &    \\
                 & $\zeta$$_{5p}$    &    2382 &   2701 & 1.134 &    \\
\hline
\end{longtable}
$^a$ F: Fixed parameter value; Rn: ratios of these parameters have been fixed in the fitting process.\\
\noindent

\begin{longtable}{llrrrl}
\caption{\label{tab:zrvii:para}Radial parameters (in cm$^{-1}$) adopted for the calculations in \ion{Zr}{vii}.} \\
\hline\hline
\noalign{\smallskip}
Configuration & Parameter & HFR & Fitted & Ratio & Note$^a$ \\ 
\noalign{\smallskip}
\hline
\endfirsthead
\caption{continued.}\\
\hline\hline
\noalign{\smallskip}
Configuration & Parameter & HFR & Fitted & Ratio & Note$^a$ \\
\noalign{\smallskip}
\hline
\noalign{\smallskip}
\endhead
\hline
\noalign{\smallskip}
\endfoot
\noalign{\smallskip}
\multicolumn{6}{c}{Even parity} \\
4p$^4$           & E$_{av}$          &   23653 &  33968 &       &    \\
                 & F$^2$(4p,4p)      &   84430 &  65839 & 0.780 &    \\
                 & $\alpha$          &       0 &    646 &       &    \\
                 & $\zeta$$_{4p}$    &   10658 &  11259 & 1.056 &    \\
4p$^3$5p         & E$_{av}$          &  516191 & 514481 &       &    \\
                 & F$^2$(4p,4p)      &   86163 &  82914 & 0.962 &    \\
                 & $\alpha$          &       0 &   -537 &       &    \\
\hline
\end{longtable}
$^a$ F: Fixed parameter value.\\
\noindent

\begin{longtable}{rrrrl}
\caption{\label{tab:zriv:ener} Comparison between available experimental and calculated energy levels in \ion{Zr}{iv}. Energies are given in cm$^{-1}$.} \\
\hline\hline
\noalign{\smallskip}
E$_{exp}$$^a$  & E$_{calc}$$^b$ & $\Delta$E & J & Leading components (in \%) in $LS$ coupling$^c$ \\ 
\noalign{\smallskip}
\hline
\endfirsthead
\caption{continued.}\\
\hline\hline
\noalign{\smallskip}
E$_{exp}$$^a$  & E$_{calc}$$^b$ & $\Delta$E & J & Leading components (in \%) in $LS$ coupling$^c$ \\
\noalign{\smallskip}
\hline
\noalign{\smallskip}
\endhead
\hline
\noalign{\smallskip}
\endfoot
\noalign{\smallskip}
\multicolumn{5}{c}{Even parity} \\
        0.00 &     0000 &   0 & 1.5 &   99  4d $^2$D  \\
     1250.70 &     1251 &   0 & 2.5 &   99  4d $^2$D  \\
    38258.35 &    38258 &   0 & 0.5 &   99  5s $^2$S  \\
   146652.40 &   146652 &   0 & 1.5 &  100  5d $^2$D  \\
   147002.46 &   147002 &   0 & 2.5 &  100  5d $^2$D  \\
   152513.00 &   152513 &   0 & 0.5 &  100  6s $^2$S  \\ 
   197765.10 &   197765 &   0 & 1.5 &  100  6d $^2$D  \\
\hline
\end{longtable}
\noindent
$^{(a)}$ From \citet{readeracquista1997}.\\
$^{(b)}$ This work.\\

\begin{longtable}{rrrrl}
\caption{\label{tab:zrv:ener} Comparison between available experimental and calculated energy levels in \ion{Zr}{v}. Energies are given in cm$^{-1}$.} \\
\hline\hline
\noalign{\smallskip}
E$_{exp}$$^a$  & E$_{calc}$$^b$ & $\Delta$E & J & Leading components (in \%) in $LS$ coupling$^c$ \\ 
\noalign{\smallskip}
\hline
\endfirsthead
\caption{continued.}\\
\hline\hline
\noalign{\smallskip}
E$_{exp}$$^a$  & E$_{calc}$$^b$ & $\Delta$E & J & Leading components (in \%) in $LS$ coupling$^c$ \\
\noalign{\smallskip}
\hline
\noalign{\smallskip}
\endhead
\hline
\noalign{\smallskip}
\endfoot
\noalign{\smallskip}
\multicolumn{5}{c}{Even parity} \\
        0.00  &       0 &      0 & 0  &  97  4p$^6$ $^1$S \\  
   371895.16  &  372099 &   -204 & 1  &  84  4p$^5$5p $^3$S     +   13  4p$^5$5p $^3$P \\    
   376897.68  &  376807 &     91 & 2  &  57  4p$^5$5p $^3$D     +   36  4p$^5$5p $^1$D     +    7  4p$^5$5p $^3$P \\   
   378753.36  &  378653 &    100 & 3  &  99  4p$^5$5p $^3$D \\    
   380855.53  &  380904 &    -48 & 1  &  46  4p$^5$5p $^1$P     +   30  4p$^5$5p $^3$D     +   20  4p$^5$5p $^3$P \\   
   382985.08  &  382952 &     33 & 2  &  67  4p$^5$5p $^3$P     +   30  4p$^5$5p $^1$D \\    
   388852.95  &  388865 &    -12 & 0  &  77  4p$^5$5p $^3$P     +   22  4p$^5$5p $^1$S \\
\hline
\end{longtable}
\noindent
$^{(a)}$ From \citet{readeracquista1979} and \citet{khanetal1981}.\\
$^{(b)}$ This work.\\
$^{(c)}$ Only the first three components that are larger than 5\% are given.

\begin{longtable}{rrrrl}
\caption{\label{tab:zrvi:ener} Comparison between available experimental and calculated energy levels in \ion{Zr}{vi}. Energies are given in cm$^{-1}$.} \\
\hline\hline
\noalign{\smallskip}
E$_{exp}$$^a$  & E$_{calc}$$^b$ & $\Delta$E & J & Leading components (in \%) in $LS$ coupling$^c$ \\ 
\noalign{\smallskip}
\hline
\endfirsthead
\caption{continued.}\\
\hline\hline
\noalign{\smallskip}
E$_{exp}$$^a$  & E$_{calc}$$^b$ & $\Delta$E & J & Leading components (in \%) in $LS$ coupling$^c$ \\
\noalign{\smallskip}
\hline
\noalign{\smallskip}
\endhead
\hline
\noalign{\smallskip}
\endfoot
\noalign{\smallskip}
\multicolumn{5}{c}{Odd parity} \\
        0.00  &       0 &      0 & 1.5 &   97  4p$^5$ $^2$P  \\  
    15602.78  &   15603 &      0 & 0.5 &   97  4p$^5$ $^2$P  \\
   421257.96  &  421364 &   -106 & 1.5 &   62  4p$^4$($^3$P)5p $^4$P  +    9  4p$^4$($^3$P)5p $^4$S  +    9  4p$^4$($^1$D)5p $^2$P  \\ 
   421991.19  &  421898 &     93 & 2.5 &   68  4p$^4$($^3$P)5p $^4$P  +   23  4p$^4$($^3$P)5p $^4$D  \\  
   425678.16  &  426017 &   -339 & 0.5 &   23  4p$^4$($^3$P)5p $^2$P  +   44  4p$^4$($^3$P)5p $^4$P  +   19  4p$^4$($^1$D)5p $^2$P  \\ 
   427118.65  &  427134 &    -15 & 2.5 &   60  4p$^4$($^3$P)5p $^2$D  +   14  4p$^4$($^3$P)5p $^4$P  +   13  4p$^4$($^3$P)5p $^4$D  \\ 
   427649.11  &  427421 &    228 & 3.5 &   89  4p$^4$($^3$P)5p $^4$D  +   10  4p$^4$($^1$D)5p $^2$F  \\
\hline
\end{longtable}
\noindent
$^{(a)}$ From  \citet{readerlindsay2016}.\\
$^{(b)}$ This work.\\
$^{(c)}$ Only the first three components that are larger than 5\% are given.

\begin{longtable}{rrrrl}
\caption{\label{tab:zrvii:ener} Comparison between available experimental and calculated energy levels in \ion{Zr}{vii}. Energies are given in cm$^{-1}$.} \\
\hline\hline
\noalign{\smallskip}
E$_{exp}$$^a$  & E$_{calc}$$^b$ & $\Delta$E & J & Leading components (in \%) in $LS$ coupling$^c$ \\ 
\noalign{\smallskip}
\hline
\endfirsthead
\caption{continued.}\\
\hline\hline
\noalign{\smallskip}
E$_{exp}$$^a$  & E$_{calc}$$^b$ & $\Delta$E & J & Leading components (in \%) in $LS$ coupling$^c$ \\
\noalign{\smallskip}
\hline
\noalign{\smallskip}
\endhead
\hline
\noalign{\smallskip}
\endfoot
\noalign{\smallskip}
\multicolumn{5}{c}{Even parity} \\
        0  &       2 &    -2 & 2 &   89  4p$^4$ $^3$P           +    9  4p$^4$ $^1$D  \\ 
    12557  &   12554 &     3 & 0 &   83  4p$^4$ $^3$P           +   14  4p$^4$ $^1$S \\ 
    13549  &   13550 &    -1 & 1 &   97  4p$^4$ $^3$P  \\ 
    27176  &   27176 &     0 & 2 &   88  4p$^4$ $^1$D           +    9  4p$^4$ $^3$P  \\ 
    56943  &   56943 &     0 & 0 &   84  4p$^4$ $^1$S           +   14  4p$^4$ $^3$P  \\  
   480659  &  480829 &  -170 & 1 &   53  4p$^3$($^4$S)5p $^3$P  +   11  4p$^3$($^4$S)5p $^5$P  +    8  4p$^3$($^2$D)5p $^3$P \\  
   483891  &  484629 &  -738 & 2 &   61  4p$^3$($^4$S)5p $^3$P  +   17  4p$^3$($^4$S)5p $^5$P  +    7  4p$^3$($^2$D)5p $^3$P \\
\hline
\end{longtable}
\noindent
$^{(a)}$ From \citet{readeracquista1976}, \citet{rahimullahetal1978}, and \citet{khanetal1983}.\\
$^{(b)}$ This work.\\
$^{(c)}$ Only the first three components that are larger than 5\% are given.

\begin{longtable}{rrccrrccrrr}
\caption{\label{tab:zriv:loggf}Calculated HFR oscillator strengths ($\log gf$) and transition probabilities ($gA$) 
                               in \ion{Zr}{iv}. CF is the
                               absolute value of the cancellation factor as defined by \citet{cowan1981}. 
                               In cols\@. 3 and 6, e is written for even and o for odd.} \\
\hline\hline
\noalign{\smallskip}
                 & \multicolumn{3}{c}{Lower level} && \multicolumn{3}{c}{Upper level} &           &      &    \\
\cline{2-4}\cline{6-8}
\noalign{\smallskip}
Wavelength{$^a$} / \AA & Energy{$^b$} / cm$^{-1}$ & Parity & $J$  && Energy{$^b$} / cm$^{-1}$ & Parity & $J$  & $\log gf$ & $gA$ / s$^{-1}$  & CF \\
\noalign{\smallskip}
\hline
\endfirsthead
\caption{continued.}\\
\hline\hline
\noalign{\smallskip}
                 & \multicolumn{3}{c}{Lower level} && \multicolumn{3}{c}{Upper level} &           &      &    \\
\cline{2-4}\cline{6-8}
\noalign{\smallskip}
Wavelength{$^a$} / \AA & Energy{$^b$} / cm$^{-1}$ & Parity & $J$  && Energy{$^b$} / cm$^{-1}$ & Parity & $J$  & $\log gf$ & $gA$ / s$^{-1}$  & CF \\
\noalign{\smallskip}
\hline
\noalign{\smallskip}
\endhead
\hline
\noalign{\smallskip}
\endfoot
\noalign{\smallskip}
     445.593 &         0 & e & 1.5 & &    224420 & o & 2.5 &  -1.61 &  $8.24 \times 10^{08} $ & 0.055 \\
     447.954 &      1251 & e & 2.5 & &    224488 & o & 3.5 &  -1.40 &  $1.33 \times 10^{09} $ & 0.066 \\
     448.091 &      1251 & e & 2.5 & &    224420 & o & 2.5 &  -2.78 &  $5.54 \times 10^{07} $ & 0.043 \\
     477.788 &         0 & e & 1.5 & &    209298 & o & 1.5 &  -2.51 &  $9.06 \times 10^{07} $ & 0.381 \\
     478.965 &         0 & e & 1.5 & &    208783 & o & 0.5 &  -1.80 &  $4.56 \times 10^{08} $ & 0.404 \\
     480.661 &      1251 & e & 2.5 & &    209298 & o & 1.5 &  -1.54 &  $8.29 \times 10^{08} $ & 0.413 \\
     497.230 &         0 & e & 1.5 & &    201114 & o & 2.5 &  -0.76 &  $4.71 \times 10^{09} $ & 0.198 \\
     500.220 &      1251 & e & 2.5 & &    201163 & o & 3.5 &  -0.59 &  $6.87 \times 10^{09} $ & 0.209 \\
\end{longtable}

\begin{longtable}{rrccrrccrrr}
\caption{\label{tab:zrv:loggf}Calculated HFR oscillator strengths ($\log gf$) and transition probabilities ($gA$) 
                               in \ion{Zr}{v}. CF is the
                               absolute value of the cancellation factor as defined by \citet{cowan1981}. 
                               In cols\@. 3 and 6, e is written for even and o for odd.} \\
\hline\hline
\noalign{\smallskip}
                 & \multicolumn{3}{c}{Lower level} && \multicolumn{3}{c}{Upper level} &           &      &    \\
\cline{2-4}\cline{6-8}
\noalign{\smallskip}
Wavelength{$^a$} / \AA & Energy{$^b$} / cm$^{-1}$ & Parity & $J$  && Energy{$^b$} / cm$^{-1}$ & Parity & $J$  & $\log gf$ & $gA$ / s$^{-1}$  & CF \\
\noalign{\smallskip}
\hline
\endfirsthead
\caption{continued.}\\
\hline\hline
\noalign{\smallskip}
                 & \multicolumn{3}{c}{Lower level} && \multicolumn{3}{c}{Upper level} &           &      &    \\
\cline{2-4}\cline{6-8}
\noalign{\smallskip}
Wavelength{$^a$} / \AA & Energy{$^b$} / cm$^{-1}$ & Parity & $J$  && Energy{$^b$} / cm$^{-1}$ & Parity & $J$  & $\log gf$ & $gA$ / s$^{-1}$  & CF \\
\noalign{\smallskip}
\hline
\noalign{\smallskip}
\endhead
\hline
\noalign{\smallskip}
\endfoot
\noalign{\smallskip}
     165.257 &         0 & e & 0.0 & &    605118 & o & 1.0 &  -2.26 &  $1.34 \times 10^{09} $ & 0.494 \\
     168.943 &         0 & e & 0.0 & &    591916 & o & 1.0 &  -2.05 &  $2.07 \times 10^{09} $ & 0.440 \\
     170.392 &         0 & e & 0.0 & &    586882 & o & 1.0 &  -2.30 &  $1.14 \times 10^{09} $ & 0.214 \\
     171.403 &         0 & e & 0.0 & &    583420 & o & 1.0 &  -2.85 &  $3.21 \times 10^{08} $ & 0.013 \\
     174.284 &         0 & e & 0.0 & &    573776 & o & 1.0 &  -6.19 &  $1.42 \times 10^{05} $ & 0.000 \\
     175.016 &         0 & e & 0.0 & &    571376 & o & 1.0 &  -2.09 &  $1.76 \times 10^{09} $ & 0.187 \\
     176.044 &         0 & e & 0.0 & &    568040 & o & 1.0 &  -2.94 &  $2.46 \times 10^{08} $ & 0.163 \\
     180.989 &         0 & e & 0.0 & &    552521 & o & 1.0 &  -1.74 &  $3.71 \times 10^{09} $ & 0.353 \\
\end{longtable}

\begin{longtable}{rrccrrccrrr}
\caption{\label{tab:zrvi:loggf}Calculated HFR oscillator strengths ($\log gf$) and transition probabilities ($gA$) 
                               in \ion{Zr}{vi}. CF is the
                               absolute value of the cancellation factor as defined by \citet{cowan1981}. 
                               In cols\@. 3 and 6, e is written for even and o for odd.} \\
\hline\hline
\noalign{\smallskip}
                 & \multicolumn{3}{c}{Lower level} && \multicolumn{3}{c}{Upper level} &           &      &    \\
\cline{2-4}\cline{6-8}
\noalign{\smallskip}
Wavelength{$^a$} / \AA & Energy{$^b$} / cm$^{-1}$ & Parity & $J$  && Energy{$^b$} / cm$^{-1}$ & Parity & $J$  & $\log gf$ & $gA$ / s$^{-1}$  & CF \\
\noalign{\smallskip}
\hline
\endfirsthead
\caption{continued.}\\
\hline\hline
\noalign{\smallskip}
                 & \multicolumn{3}{c}{Lower level} && \multicolumn{3}{c}{Upper level} &           &      &    \\
\cline{2-4}\cline{6-8}
\noalign{\smallskip}
Wavelength{$^a$} / \AA & Energy{$^b$} / cm$^{-1}$ & Parity & $J$  && Energy{$^b$} / cm$^{-1}$ & Parity & $J$  & $\log gf$ & $gA$ / s$^{-1}$  & CF \\
\noalign{\smallskip}
\hline
\noalign{\smallskip}
\endhead
\hline
\noalign{\smallskip}
\endfoot
\noalign{\smallskip}
     165.931 &         0 & o & 1.5 & &    602661 & e & 0.5 &  -2.02 &  $2.32 \times 10^{09} $ & 0.098 \\
     170.341 &     15603 & o & 0.5 & &    602661 & e & 0.5 &  -1.48 &  $7.65 \times 10^{09} $ & 0.467 \\
     173.947 &         0 & o & 1.5 & &    574889 & e & 1.5 &  -3.12 &  $1.68 \times 10^{08} $ & 0.005 \\
     174.066 &         0 & o & 1.5 & &    574495 & e & 2.5 &  -0.89 &  $2.81 \times 10^{10} $ & 0.515 \\
     174.428 &         0 & o & 1.5 & &    573301 & e & 1.5 &  -3.83 &  $3.25 \times 10^{07} $ & 0.002 \\
     174.489 &         0 & o & 1.5 & &    573102 & e & 2.5 &  -2.49 &  $7.02 \times 10^{08} $ & 0.037 \\
     178.237 &         0 & o & 1.5 & &    561050 & e & 0.5 &  -1.25 &  $1.19 \times 10^{10} $ & 0.483 \\
     178.777 &         0 & o & 1.5 & &    559356 & e & 1.5 &  -1.26 &  $1.13 \times 10^{10} $ & 0.512 \\
\end{longtable}

\begin{longtable}{rrccrrccrrr}
\caption{\label{tab:zrvii:loggf}Calculated HFR oscillator strengths ($\log gf$) and transition probabilities ($gA$) 
                               in \ion{Zr}{vii}. CF is the
                               absolute value of the cancellation factor as defined by \citet{cowan1981}. 
                               In cols\@. 3 and 6, e is written for even and o for odd.} \\
\hline\hline
\noalign{\smallskip}
                 & \multicolumn{3}{c}{Lower level} && \multicolumn{3}{c}{Upper level} &           &      &    \\
\cline{2-4}\cline{6-8}
\noalign{\smallskip}
Wavelength{$^a$} / \AA & Energy{$^b$} / cm$^{-1}$ & Parity & $J$  && Energy{$^b$} / cm$^{-1}$ & Parity & $J$  & $\log gf$ & $gA$ / s$^{-1}$  & CF \\
\noalign{\smallskip}
\hline
\endfirsthead
\caption{continued.}\\
\hline\hline
\noalign{\smallskip}
                 & \multicolumn{3}{c}{Lower level} && \multicolumn{3}{c}{Upper level} &           &      &    \\
\cline{2-4}\cline{6-8}
\noalign{\smallskip}
Wavelength{$^a$} / \AA & Energy{$^b$} / cm$^{-1}$ & Parity & $J$  && Energy{$^b$} / cm$^{-1}$ & Parity & $J$  & $\log gf$ & $gA$ / s$^{-1}$  & CF \\
\noalign{\smallskip}
\hline
\noalign{\smallskip}
\endhead
\hline
\noalign{\smallskip}
\endfoot
\noalign{\smallskip}
     213.117 &         0 & e & 2.0 & &    469225 & o & 1.0 &  -2.32 &  $7.09 \times 10^{08} $ & 0.010 \\
     214.536 &         0 & e & 2.0 & &    466123 & o & 2.0 &  -1.18 &  $9.67 \times 10^{09} $ & 0.060 \\
     218.320 &         0 & e & 2.0 & &    458043 & o & 1.0 &  -1.64 &  $3.18 \times 10^{09} $ & 0.054 \\
     218.977 &     12557 & e & 0.0 & &    469225 & o & 1.0 &  -1.92 &  $1.67 \times 10^{09} $ & 0.024 \\
     219.454 &     13549 & e & 1.0 & &    469225 & o & 1.0 &  -1.90 &  $1.75 \times 10^{09} $ & 0.050 \\
     220.958 &     13549 & e & 1.0 & &    466123 & o & 2.0 &  -0.52 &  $4.12 \times 10^{10} $ & 0.594 \\
     224.474 &     12557 & e & 0.0 & &    458043 & o & 1.0 &  -0.66 &  $2.89 \times 10^{10} $ & 0.597 \\
     224.975 &     13549 & e & 1.0 & &    458043 & o & 1.0 &  -1.23 &  $7.86 \times 10^{09} $ & 0.287 \\
\end{longtable}

\section{Additional tables for xenon}
\label{app:addtabxe}

\begin{longtable}{llrrrl}
\caption{\label{tab:xeiv:para}Radial parameters (in cm$^{-1}$) adopted for the calculations in \ion{Xe}{iv}.} \\
\hline\hline
\noalign{\smallskip}
Configuration & Parameter & HFR & Fitted & Ratio & Note$^a$ \\ 
\noalign{\smallskip}
\hline
\endfirsthead
\caption{continued.}\\
\hline\hline
\noalign{\smallskip}
Configuration & Parameter & HFR & Fitted & Ratio & Note$^a$ \\
\noalign{\smallskip}
\hline
\noalign{\smallskip}
\endhead
\hline
\noalign{\smallskip}
\endfoot
\noalign{\smallskip}
\multicolumn{6}{c}{Odd parity} \\
5p$^3$           & E$_{av}$          &   29571 &  29733 &       &    \\
                 & F$^2$(5p,5p)      &   53594 &  48082 & 0.897 &    \\    
                 & $\alpha$          &       0 &   -105 &       &    \\
                 & $\zeta$$_{5p}$    &    8331 &   9017 & 1.082 &    \\
5p$^2$6p         & E$_{av}$          &  220434 & 210991 &       &    \\
                 & F$^2$(5p,5p)      &   55341 &  42311 & 0.764 &    \\  
                 & $\alpha$          &       0 &   -136 &       &    \\
\hline
\end{longtable}
$^a$ F: Fixed parameter value; R: ratios of these parameters had been fixed in the fitting process.\\
\noindent

\begin{longtable}{llrrrl}
\caption{\label{tab:xev:para}Radial parameters (in cm$^{-1}$) adopted for the calculations in \ion{Xe}{v}.} \\
\hline\hline
\noalign{\smallskip}
Configuration & Parameter & HFR & Fitted & Ratio & Note$^a$ \\ 
\noalign{\smallskip}
\hline
\endfirsthead
\caption{continued.}\\
\hline\hline
\noalign{\smallskip}
Configuration & Parameter & HFR & Fitted & Ratio & Note$^a$ \\
\noalign{\smallskip}
\hline
\noalign{\smallskip}
\endhead
\hline
\noalign{\smallskip}
\endfoot
\noalign{\smallskip}
\multicolumn{6}{c}{Even parity} \\
5p$^2$           & E$_{av}$          &   28481 &  29415 &       &    \\
                 & F$^2$(5p,5p)      &   55631 &  50605 & 0.910 &    \\    
                 & $\alpha$          &       0 &   -111 &       &    \\
                 & $\zeta$$_{5p}$    &    9121 &   9730 & 1.067 &    \\
5p6p             & E$_{av}$          &  253537 & 252141 &       &    \\
                 & $\zeta$$_{5p}$    &    9817 &   8816 & 0.898 &    \\
                 & $\zeta$$_{6p}$    &    2614 &   2694 & 1.031 &    \\
\hline
\end{longtable}
$^a$ F: Fixed parameter value; R: ratios of these parameters have been fixed in the fitting process.\\
\noindent

\begin{longtable}{rrrrl}
\caption{\label{tab:xeiv:ener} Comparison between available experimental and calculated energy levels in \ion{Xe}{iv}. Energies are given in cm$^{-1}$.} \\
\hline\hline
\noalign{\smallskip}
E$_{exp}$$^a$  & E$_{calc}$$^b$ & $\Delta$E & J & Leading components (in \%) in $LS$ coupling$^c$ \\ 
\noalign{\smallskip}
\hline
\endfirsthead
\caption{continued.}\\
\hline\hline
\noalign{\smallskip}
E$_{exp}$$^a$  & E$_{calc}$$^b$ & $\Delta$E & J & Leading components (in \%) in $LS$ coupling$^c$ \\
\noalign{\smallskip}
\hline
\noalign{\smallskip}
\endhead
\hline
\noalign{\smallskip}
\endfoot
\noalign{\smallskip}
\multicolumn{5}{c}{Odd parity} \\

        0.0 &        3 &     -3 & 1.5 &   81  5p$^3$ $^4$S           +  12  5p$^3$ $^2$P \\   
    13267.0 &    13248 &     19 & 1.5 &   72  5p$^3$ $^2$D           +  13  5p$^3$ $^2$P           +  11  5p$^3$ $^4$S \\  
    17510.7 &    17524 &    -13 & 2.5 &   96  5p$^3$ $^2$D \\   
    28036.4 &    28043 &     -7 & 0.5 &   96  5p$^3$ $^2$P \\ 
    35649.6 &    35644 &      5 & 1.5 &   70  5p$^3$ $^2$P           +  20  5p$^3$ $^2$D           +   5  5p$^3$ $^4$S \\  
   180151.5 &   180062 &     90 & 2.5 &   68  5p$^2$($^3$P)4f $^4$G  +   7  5p$^2$($^1$S)4f $^2$F  +   7  5p$^2$($^3$P)4f $^4$F \\  
   182219.1 &   182422 &   -203 & 3.5 &   50  5p$^2$($^3$P)4f $^4$G  +  14  5p$^2$($^3$P)4f $^4$D  +  14  5p$^2$($^3$P)4f $^4$F \\
\hline
\end{longtable}
\noindent
$^{(a)}$ From  \citet{saloman2004}.\\
$^{(b)}$ This work.\\
$^{(c)}$ Only the first three components that are larger than 5\% are given.

\begin{longtable}{rrrrl}
\caption{\label{tab:xev:ener} Comparison between available experimental and calculated energy levels in \ion{Xe}{v}. Energies are given in cm$^{-1}$.} \\
\hline\hline
\noalign{\smallskip}
E$_{exp}$$^a$  & E$_{calc}$$^b$ & $\Delta$E & J & Leading components (in \%) in $LS$ coupling$^c$ \\ 
\noalign{\smallskip}
\hline
\endfirsthead
\caption{continued.}\\
\hline\hline
\noalign{\smallskip}
E$_{exp}$$^a$  & E$_{calc}$$^b$ & $\Delta$E & J & Leading components (in \%) in $LS$ coupling$^c$ \\
\noalign{\smallskip}
\hline
\noalign{\smallskip}
\endhead
\hline
\noalign{\smallskip}
\endfoot
\noalign{\smallskip}
\multicolumn{5}{c}{Even parity} \\
        0.0 &       -6 &      6 & 0 &   87  5p$^2$  $^3$P    +  10  5p$^2$    $^1$S \\  
     9291.8 &     9290 &      2 & 1 &   97  5p$^2$  $^3$P \\    
    14126.7 &    14142 &    -15 & 2 &   66  5p$^2$  $^3$P    +  31  5p$^2$    $^1$D \\    
    28411.2 &    28402 &      9 & 2 &   65  5p$^2$  $^1$D    +  31  5p$^2$    $^3$P \\    
    44470.4 &    44471 &     -1 & 0 &   86  5p$^2$  $^1$S    +  10  5p$^2$    $^3$P \\    
   186746.7 &   186635 &    111 & 3 &   54  5p4f $^3$G       +  37  5p4f   $^1$F \\    
   189663.8 &   189859 &   -196 & 3 &   44  5p4f $^3$F       +  21  5p4f   $^3$D       + 21  5p4f   $^1$F \\
\hline
\end{longtable}
\noindent
$^{(a)}$ From  \citet{saloman2004} and  \citet{rainerietal2009}.\\
$^{(b)}$ This work.\\
$^{(c)}$ Only the first three components that are larger than 5\% are given.

\begin{longtable}{rrccrrccrrr}
\caption{\label{tab:xeiv:loggf}Calculated HFR oscillator strengths ($\log gf$) and transition probabilities ($gA$) 
                               in \ion{Xe}{iv}. CF is the
                               absolute value of the cancellation factor as defined by \citet{cowan1981}. 
                               In cols\@. 3 and 6, e is written for even and o for odd.} \\
\hline\hline
\noalign{\smallskip}
                 & \multicolumn{3}{c}{Lower level} && \multicolumn{3}{c}{Upper level} &           &      &    \\
\cline{2-4}\cline{6-8}
\noalign{\smallskip}
Wavelength{$^a$} / \AA & Energy{$^b$} / cm$^{-1}$ & Parity & $J$  && Energy{$^b$} / cm$^{-1}$ & Parity & $J$  & $\log gf$ & $gA$ / s$^{-1}$  & CF \\
\noalign{\smallskip}
\hline
\endfirsthead
\caption{continued.}\\
\hline\hline
\noalign{\smallskip}
                 & \multicolumn{3}{c}{Lower level} && \multicolumn{3}{c}{Upper level} &           &      &    \\
\cline{2-4}\cline{6-8}
\noalign{\smallskip}
Wavelength{$^a$} / \AA & Energy{$^b$} / cm$^{-1}$ & Parity & $J$  && Energy{$^b$} / cm$^{-1}$ & Parity & $J$  & $\log gf$ & $gA$ / s$^{-1}$  & CF \\
\noalign{\smallskip}
\hline
\noalign{\smallskip}
\endhead
\hline
\noalign{\smallskip}
\endfoot
\noalign{\smallskip}
     494.916 &         0 & o & 1.5 & &    202055 & e & 0.5 &  -4.80 &  $4.29 \times 10^{05} $ & 0.000 \\
     525.295 &         0 & o & 1.5 & &    190369 & e & 1.5 &  -2.53 &  $7.24 \times 10^{07} $ & 0.003 \\
     526.231 &         0 & o & 1.5 & &    190031 & e & 2.5 &  -1.79 &  $3.87 \times 10^{08} $ & 0.003 \\
     529.696 &     13267 & o & 1.5 & &    202055 & e & 0.5 &  -3.32 &  $1.14 \times 10^{07} $ & 0.001 \\
     531.145 &         0 & o & 1.5 & &    188273 & e & 0.5 &  -2.84 &  $3.43 \times 10^{07} $ & 0.001 \\
     533.200 &         0 & o & 1.5 & &    187547 & e & 1.5 &  -2.03 &  $2.18 \times 10^{08} $ & 0.003 \\
     537.494 &         0 & o & 1.5 & &    186049 & e & 2.5 &  -1.74 &  $4.20 \times 10^{08} $ & 0.011 \\
     547.732 &         0 & o & 1.5 & &    182571 & e & 1.5 &  -1.79 &  $3.64 \times 10^{08} $ & 0.004 \\
\end{longtable}

\begin{longtable}{rrccrrccrrr}
\caption{\label{tab:xev:loggf}Calculated HFR oscillator strengths ($\log gf$) and transition probabilities ($gA$) 
                               in \ion{Xe}{v}. CF is the
                               absolute value of the cancellation factor as defined by \citet{cowan1981}. 
                               In cols\@. 3 and 6, e is written for even and o for odd.} \\
\hline\hline
\noalign{\smallskip}
                 & \multicolumn{3}{c}{Lower level} && \multicolumn{3}{c}{Upper level} &           &      &    \\
\cline{2-4}\cline{6-8}
\noalign{\smallskip}
Wavelength{$^a$} / \AA & Energy{$^b$} / cm$^{-1}$ & Parity & $J$  && Energy{$^b$} / cm$^{-1}$ & Parity & $J$  & $\log gf$ & $gA$ / s$^{-1}$  & CF \\
\noalign{\smallskip}
\hline
\endfirsthead
\caption{continued.}\\
\hline\hline
\noalign{\smallskip}
                 & \multicolumn{3}{c}{Lower level} && \multicolumn{3}{c}{Upper level} &           &      &    \\
\cline{2-4}\cline{6-8}
\noalign{\smallskip}
Wavelength{$^a$} / \AA & Energy{$^b$} / cm$^{-1}$ & Parity & $J$  && Energy{$^b$} / cm$^{-1}$ & Parity & $J$  & $\log gf$ & $gA$ / s$^{-1}$  & CF \\
\noalign{\smallskip}
\hline
\noalign{\smallskip}
\endhead
\hline
\noalign{\smallskip}
\endfoot
\noalign{\smallskip}
     318.590 &         0 & e & 0.0 & &    313883 & o & 1.0 &  -2.56 &  $1.82 \times 10^{08} $ & 0.036 \\
     326.728 &         0 & e & 0.0 & &    306065 & o & 1.0 &  -3.87 &  $8.40 \times 10^{06} $ & 0.001 \\
     328.309 &      9292 & e & 1.0 & &    313883 & o & 1.0 &  -1.88 &  $8.24 \times 10^{08} $ & 0.557 \\
     329.311 &      9292 & e & 1.0 & &    312956 & o & 2.0 &  -1.10 &  $4.88 \times 10^{09} $ & 0.644 \\
     331.614 &         0 & e & 0.0 & &    301555 & o & 1.0 &  -2.78 &  $1.02 \times 10^{08} $ & 0.007 \\
     333.604 &     14127 & e & 2.0 & &    313883 & o & 1.0 &  -1.61 &  $1.46 \times 10^{09} $ & 0.100 \\
     334.639 &     14127 & e & 2.0 & &    312956 & o & 2.0 &  -0.81 &  $9.18 \times 10^{09} $ & 0.634 \\
     335.511 &         0 & e & 0.0 & &    298053 & o & 1.0 &  -1.23 &  $3.48 \times 10^{09} $ & 0.482 \\
\end{longtable}

\begin{longtable}{rrccrrccrrr}
\caption{\label{tab:xevii:loggf}Calculated HFR oscillator strengths ($\log gf$) and transition probabilities ($gA$) 
                               in \ion{Xe}{vii}. CF is the
                               absolute value of the cancellation factor as defined by \citet{cowan1981}. 
                               In cols\@. 3 and 6, e is written for even and o for odd.} \\
\hline\hline
\noalign{\smallskip}
                 & \multicolumn{3}{c}{Lower level} && \multicolumn{3}{c}{Upper level} &           &      &    \\
\cline{2-4}\cline{6-8}
\noalign{\smallskip}
Wavelength{$^a$} / \AA & Energy{$^b$} / cm$^{-1}$ & Parity & $J$  && Energy{$^b$} / cm$^{-1}$ & Parity & $J$  & $\log gf$ & $gA$ / s$^{-1}$  & CF \\
\noalign{\smallskip}
\hline
\endfirsthead
\caption{continued.}\\
\hline\hline
\noalign{\smallskip}
                 & \multicolumn{3}{c}{Lower level} && \multicolumn{3}{c}{Upper level} &           &      &    \\
\cline{2-4}\cline{6-8}
\noalign{\smallskip}
Wavelength{$^a$} / \AA & Energy{$^b$} / cm$^{-1}$ & Parity & $J$  && Energy{$^b$} / cm$^{-1}$ & Parity & $J$  & $\log gf$ & $gA$ / s$^{-1}$  & CF \\
\noalign{\smallskip}
\hline
\noalign{\smallskip}
\endhead
\hline
\noalign{\smallskip}
\endfoot
\noalign{\smallskip}
     204.100 &         0 & e & 0.0 & &    489957 & o & 1.0 &  -3.88 &  $2.11 \times 10^{07} $ & 0.001 \\
     212.402 &         0 & e & 0.0 & &    470805 & o & 1.0 &  -4.20 &  $9.33 \times 10^{06} $ & 0.002 \\
     226.564 &         0 & e & 0.0 & &    441376 & o & 1.0 &  -2.58 &  $3.41 \times 10^{08} $ & 0.005 \\
     235.534 &         0 & e & 0.0 & &    424567 & o & 1.0 &  -3.76 &  $2.10 \times 10^{07} $ & 0.037 \\
     243.296 &         0 & e & 0.0 & &    411022 & o & 1.0 &  -1.90 &  $1.42 \times 10^{09} $ & 0.108 \\
     245.217 &         0 & e & 0.0 & &    407802 & o & 1.0 &  -1.07 &  $9.39 \times 10^{09} $ & 0.422 \\
     249.443 &         0 & e & 0.0 & &    400893 & o & 1.0 &  -1.52 &  $3.26 \times 10^{09} $ & 0.444 \\
     263.263 &     96141 & o & 0.0 & &    475990 & e & 1.0 &  -3.14 &  $7.02 \times 10^{07} $ & 0.043 \\
\end{longtable}

\twocolumn
\end{appendix}

\end{document}